\documentclass[epj]{revtex4}

\usepackage{graphicx}
\usepackage{amssymb}
\usepackage{amsmath}

\begin{document}

\title{\Large \textbf{Quantum entanglement in exactly soluble atomic models:
The Moshinsky model with three electrons, and with two electrons in a uniform magnetic field.}}
\author{P.A. Bouvrie$^{a}$, A.P. Majtey$^{a}$, A.R. Plastino$^{a,b}$, P. S\'anchez-Moreno$^{a,c}$, J.S. Dehesa$^{a}$}
\affiliation{
$^a$Instituto {\em Carlos I} de F\'{\i}sica Te\'orica y Computacional and
Departamento de F\'{\i}sica At\'omica, Molecular y Nuclear, Universidad de Granada, 18071-Granada, Spain\\
$^b$Universidad Nacional de La Plata, CREG-UNLP, C.C. 727, 1900 La Plata, Argentina\\
$^c$Departamento de Matem\'atica Aplicada, Universidad de Granada, 18071-Granada, Spain\\
$^{*}$(corresponding author arplastino@ugr.es)
}

\date{\today}

\begin{abstract}
We investigate the entanglement-related features of the eigenstates of two exactly soluble atomic models: a one-dimensional three-electron Moshinsky model, and a three-dimensional two-electron Moshinsky system in an external uniform magnetic field. We analytically compute the amount of entanglement exhibited by the wavefunctions corresponding to the ground, first and second excited states of the three-electron model. We found that the amount of entanglement of the system tends to increase with energy, and in the case of excited states we found a finite amount of entanglement in the limit of vanishing interaction. We also analyze the entanglement properties of the ground and first few excited states of the two-electron Moshinsky model in the presence of a magnetic field. The dependence of the eigenstates' entanglement on the energy, as well as its behaviour in the regime of vanishing interaction, are similar to those observed in the three-electron system. On the other hand, th
 e entanglement exhibits a monotonically decreasing behavior with the strength of the external magnetic field. For strong magnetic fields the entanglement approaches a finite asymptotic value that depends on the interaction strength. For both systems studied here we consider a perturbative approach in order to shed some light on the entanglement's dependence on energy and also to clarify the finite entanglement exhibited by excited states in the limit of weak interactions. As far as we know, this is the first work that provides analytical and exact results for the entanglement properties of a three-electron model.
\end{abstract}

\maketitle

\section{Introduction}

Entanglement is an essential ingredient of the quantum mechanical description of Nature \cite{AFOV08,TMB10,BEN}. Besides its central role for the basic understanding of
the quantum world, entanglement constitutes a physical resource admitting numerous technological applications. The study of entanglement sheds new light on the mechanisms behind the quantum-to-classical transition \cite{SCH} as well as on the foundations of statistical mechanics \cite{GMM}. On the other hand, the controlled manipulation of entangled states of multipartite systems is fundamental for the implementation of quantum information processes, such as quantum computation \cite{NIE,ABF}.
Quantum entanglement is also relevant in connection with the physical characterization of atoms and molecules. The exploration of the entanglement features exhibited by  atoms and molecules is a captivating field of enquiry because these composite quantum objects play a central role in our understanding of both Nature and technology.
In point of fact, the entanglement properties of atomic systems have been the subject of considerable research activity in recent years \cite{YPD,OSE,OS08,AM1,TMB10,CMS,CSD08,PN09,MPDK10,HF11,K11}. This line of research is contained within the more general one aimed at the application of information-theoretic concepts and methods to the study of atomic and molecular systems \cite{DAS,GON,LAAE,NC1,NC2,DGS,PLA,SAA,NS1,L07,GSE03,GN05,LS2011}.

Some of the most detailed results on the entanglement properties of atomic systems, particularly in the case of excited states, have been obtained from
analytical investigations of soluble two-electron models \cite{YPD,MPDK10}. Partial results were also obtained numerically for the eigenstates of helium-like
systems, employing high quality wave functions \cite{MPDK10}.  Some general trends are beginning to emerge from these investigations. It is observed that the amount of entanglement of the atomic eigenstates tends to increase with the concomitant energy. It  also increases with the strength of the interaction between the constituent particles. On the other hand, the entanglement of excited states shows an apparent discontinuous behaviour: it does not necessarily vanish in the limit of very small interactions \cite{YPD}. It would be desirable to extend these studies to more general scenarios, particularly to models consisting of more than
two electrons, or involving magnetic fields. The aim of the present contribution is to investigate the entanglement properties of the eigenstates of the
exactly soluble Moshinsky model \cite{MO1}, extending previous works to the cases of a three-electron system and a three-dimensional two-electron system in a uniform
external magnetic field.

The paper is organized as follows. In section 2 we briefly discuss entanglement in systems of identical fermions.
We review the measure used in order of quantify the amount of entanglement of pure states, focusing on
appropriate measures for two- and three-electrons systems. In section 3 we investigate the entanglement properties of
the eigenstates of the Moshinsky model with three electrons.  The entanglement features of the three-dimensional
Moshinsky model with two electrons in the presence of a uniform magnetic field are studied in section 4. Then,
in section 5 we consider a perturbative approach to clarify some entanglement features found in the previous models.
Finally, some conclusions are drawn in section 6.

\section{Entanglement measure}

Correlations between two identical fermions that are only due to
the antisymmetric nature of the two-particle state do not
contribute to the state's entanglement
\cite{ESB,GH1,GH2,NAU,BPC,OST}. The entanglement of
the two-fermion state is given by the quantum correlations
existing on top of these minimum ones. A practical quantitative
measure for the amount of entanglement exhibited by  a pure state
$|\psi\rangle$ of a system of $N$ identical fermions is (see
\cite{PMD09} and references therein) that given (up to an appropriate
multiplicative and additive constant) by the linear entropy
of the single particle reduced density matrix  $\rho_r$,
\begin{equation} \label{entanglementSL}
\varepsilon(|\psi\rangle)=1-NTr[\rho_r^2],
\end{equation}
Notice that, according to this entanglement measure, a pure
state that takes the form of a single Slater determinant has
no entanglement. The measure (\ref{entanglementSL}) is normalized
to adopt values in the interval $[0,1]$.

We shall apply the measure given by Eq. \eqref{entanglementSL} to a pure state $|\Psi\rangle$
of a one dimensional system consisting of three spin-$\frac{1}{2}$ fermions (electrons).
This pure state has an associated wave function given, in self-explanatory notation, by
$\Psi(x_1 \sigma_1,x_2 \sigma_2,x_3 \sigma_3) = \langle x_1  \sigma_1 ,x_2 \sigma_2,x_3 \sigma_3|
 \Psi\rangle$ with  $|x_1  \sigma_1 ,x_2 \sigma_2,x_3 \sigma_3 \rangle =
| x_1,x_2,x_3\rangle \otimes | \sigma_1,\sigma_2, \sigma_3 \rangle$.
 Here $x_{1,2,3}$ are the coordinates of the three electrons and the dichotomic
variables $\sigma_{1,2,3}$ (each adopting the possible values $\pm$ and corresponding to
the $S_z$ component of spin) describe  the spin degrees of freedom of the three electrons.
In order to evaluate the amount of entanglement of the system we have to compute the
following integrals,

\begin{equation}
\langle x_1\sigma_1|\rho_r|x'_1\sigma'_1\rangle = \sum_{\sigma_2,\sigma_3=\pm} \int_{-\infty}^{\infty} \langle x_1\sigma_1,x_2\sigma_2,x_3\sigma_3|\rho|x'_1\sigma'_1,x_2\sigma_2,x_3\sigma_3\rangle dx_2 dx_3
\end{equation}
where $\langle x_1\sigma_1|\rho_r|x'_1\sigma'_1\rangle$ are the elements of the one-particle reduced density matrix $\rho_r(x_1,x_1')$, $\rho=|\psi\rangle \langle \psi|$ and
\[
 \langle x_1\sigma_1,x_2\sigma_2,x_3\sigma_3|\rho|x'_1\sigma'_1,x_2\sigma_2,x_3\sigma_3\rangle = \Psi(x_1\sigma_1,x_2\sigma_2,x_3\sigma_3)\Psi^*(x'_1\sigma'_1,x_2\sigma_2,x_3\sigma_3).
\]
The square reduced spin density matrix is given by
\[
 \langle x_1\sigma_1|\rho^2_r|x'_1\sigma'_1\rangle = \sum_{\sigma=\pm} \int_{-\infty}^{\infty} \langle x_1\sigma_1|\rho_r|x \sigma \rangle \langle x \sigma|\rho_r |x'_1\sigma'_1 \rangle dx
\]
and finally the expression for the trace is
\[
 Tr[\rho^2_r] = \sum_{\sigma=\pm } \int_{-\infty}^{\infty} \langle x\sigma|\rho^2_r|x \sigma \rangle dx
\]

In the three-electron case it is not possible to find totally antisymmetric factorizable between coordinates and spin wave functions, however in the two-electron case, following \cite{YPD} we focus on states with factorized wave functions. The corresponding density matrix takes the form

\begin{equation}
 \rho=\rho^{(c)}\otimes\rho^{(s)}
\end{equation}
and then, the entanglement measure evaluated on these states is given by
\begin{equation}
 \varepsilon=1-2Tr[(\rho_r^{(c)})^2]Tr[(\rho_r^{(s)})^2],
\end{equation}
where $\rho_r^{(c)}$ and $\rho_r^{(s)}$ are the single-particle reduced coordinate and spin density matrices. So, in the case of two-electron system studied in section 4, we consider separately the cases of parallel and antiparallel spin wave function. In the case of parallel spins, described by $|++\rangle$ or $|--\rangle$, the coordinate wave function must be antisymmetric and we have $Tr[(\rho_r^{(s)})^2]=1$. On the other hand if we have antiparallel spins, we can distinguish two cases: symmetric coordinate wave function with spin wave function of the form $\frac{1}{\sqrt{2}}(|+-\rangle-|-+\rangle)$ or  antisymmetric coordinate wave function with spin wave function $\frac{1}{\sqrt{2}}(|+-\rangle+|-+\rangle)$, both of them with $Tr[(\rho_r^{(s)})^2]=\frac{1}{2}$. And finally, to calculate the amount of entanglement we will compute the integrals
\begin{equation}
 \langle \mathbf{r}_1|\rho_r^{(c)}|\mathbf{r'}_1 \rangle = \int_{\mathbb{R}^3} \langle \mathbf{r}_1\mathbf{r}_2|\rho^{(c)}|\mathbf{r'}_1 \mathbf{r}_2 \rangle d\mathbf{r}_2 = \int_{\mathbb{R}^3} \Psi(\mathbf{r}_1,\mathbf{r}_2)\Psi^*(\mathbf{r'}_1,\mathbf{r}_2) d\mathbf{r}_2
\end{equation}
and the trace of the coordinate part is
\begin{equation}
Tr[(\rho_r^{(c)})^2]=\int_{\mathbb{R}^3} |\langle \mathbf{r}_1|\rho_r^{(c)}|\mathbf{r'}_1\rangle|^2 d\mathbf{r}_1d\mathbf{r'}_1
\end{equation}

\section{The three-electron Moshinsky atom}

The Moshinsky atom \cite{MO1} is a system formed by harmonically interacting particles confined in a common, external isotropic harmonic potential. The total Hamiltonian of the one-dimensional Moshinsky atom with three electrons is
\begin{equation} \label{hami}
 H=-\frac{1}{2}\left( \frac{\partial^2}{\partial x_1^2}+\frac{\partial^2}{\partial x_2^2}+\frac{\partial^2}{\partial x_3^2} \right) + \frac{1}{2} \omega^2 (x_1^2+x_2^2+x_3^2)\pm\frac{1}{2}\lambda^2 [(x_1-x_2)^2+(x_2-x_3)^2+(x_3-x_1)^2]
\end{equation}
where $x_1$, $x_2$ and $x_3$ are the coordinates of the three particles, $\omega$ is the natural frequency of the external harmonic field, and $\lambda$ is the natural frequency of the interaction harmonic field. The positive sign in the last term describes an attractive interaction between the electrons and the negative a repulsive interaction. We use atomic units ($m_e=1$, $\hbar=1$) throughout the paper, unless indicated otherwise.

Introducing the Jacobi coordinates for three particles,
\begin{equation}
 R_1=\frac{1}{\sqrt{3}}(x_1+x_2+x_3), \, R_2=\frac{1}{\sqrt{6}}(-2x_1+x_2+x_3) \text{ \ \ and \ \ } R_3=\frac{1}{\sqrt{2}}(x_2-x_3)
\end{equation}
the Hamiltonian separates in the following way,
\begin{equation}
\label{Hamiltonian}
H=\left(-\frac{1}{2}\frac{\partial^2}{\partial R_1^2} + \frac{1}{2} \beta_1 R_1^2 \right) +   \left(-\frac{1}{2}\frac{\partial^2}{\partial R_2^2} + \frac{1}{2} \beta_2 R_2^2 \right)+\left(-\frac{1}{2}\frac{\partial^2}{\partial R_3^2} + \frac{1}{2} \beta_3 R_3^2 \right)
\end{equation}
where $\beta_1=\omega^2$ and
$\beta_2=\beta_3=\Lambda^2=\omega^2\pm3\lambda^2$ (again, the $+$
sign corresponds to an attractive interaction, while the $-$ sign
corresponds to a repulsive one). In the case of a repulsive
interaction it is necessary to impose the constraint $\lambda <
\frac{\omega}{\sqrt{3}}$ in order to obtain bound eigenstates.
The general eigenfunctions of the system are
\begin{equation}
\label{wavefunction}
 \Psi(x_1,x_2,x_3)=\Psi(R_1,R_2,R_3)=\Psi_{n_{R_1}}(R_1)\Psi_{n_{R_2}}(R_2)\Psi_{n_{R_3}}(R_3)
\end{equation}
with
\begin{equation}
\Psi_{n_{R_i}}(R_i)= \left( \frac{\beta_i^{1/4}}{2^{n_{R_i}}n_{R_i}!\pi^{1/2}} \right)^\frac{1}{2} e^{-\frac{1}{2}\sqrt{\beta_i}R_i^2} H_{n_{R_i}}\left( \beta_i^{1/4}R_i \right),
\end{equation}
where $H_n(x)$ denote the Hermite polynomials. The eigenenergies of these states are
\begin{equation}
 E=E_{R_1}+E_{R_2}+E_{R_3}=\omega\left(n_{R_1}+\frac{1}{2}\right)+\Lambda\left(n_{R_2}+n_{R_3}+1\right)
\end{equation}

We will denote by $|n_{R_1}n_{R_2}n_{R_3}\rangle$ the eigenstates of the Hamiltonian \eqref{Hamiltonian},
which are characterized by the three quantum numbers $n_{R_1}$, $n_{R_2}$ and $n_{R_3}$.
To fully define the three-electron system's eigenstates we must take into account combinations of
such functions of the coordinates together with the spin ones $|\sigma_1\sigma_2\sigma_3\rangle$ to
obtain total antisymmetric wave functions. In this case the wave functions corresponding to the
energy eigenstates cannot always be chosen to be separable between coordinates and spin and there
are no spin functions totally antisymmetric by themselves. The Hamiltonian commutes with the spin observables,
since it does not explicitly involve the spins. In particular, it commutes with the total $z$-component
of spin angular momentum $S_z$. Consequently, it is possible to choose energy eigenstates that are also
eigenstates of $S_z$. It is plain that the wave functions associated with these eigenstates can always be
written (up to a global normalization constant) in one of the forms
\begin{equation} \label{forabe1}
|\Phi^{(++-)}\rangle |++-\rangle \, + \,
|\Phi^{(+-+)}\rangle |+-+\rangle \, + \,
|\Phi^{(-++)}\rangle |-++\rangle,
\end{equation}
\begin{equation} \label{forabe2}
|\Phi^{(+++)}\rangle |+++\rangle,
\end{equation}
or in the forms obtained substituting $+$ by $-$ (and viceversa) in the above expressions.
In (\ref{forabe1}-\ref{forabe2}) the kets $|\Phi^{(\sigma_1, \sigma_2, \sigma_3)}\rangle$
correspond to the translational degrees of freedom and have associated coordinate
wave functions $\Phi^{(\sigma_1, \sigma_2, \sigma_3)}(x_1,x_2,x_3)= \langle x_1, x_2, x_3 |
\Phi^{(\sigma_1, \sigma_2, \sigma_3)}\rangle$. For the states (\ref{forabe1}-\ref{forabe2})
to be fully antisymmetric the coordinate wave functions  $\Phi^{(\sigma_1, \sigma_2, \sigma_3)}(x_1,x_2,x_3)$
must satisfy the following set of relations. If $\sigma_1 = \sigma_2$ we must have
$\Phi^{(\sigma_1, \sigma_2, \sigma_3)}(x_2,x_1,x_3) = - \Phi^{(\sigma_1, \sigma_2, \sigma_3)}(x_1,x_2,x_3)$,
(that is, in this case the coordinate wave function $\Phi^{(\sigma_1, \sigma_2, \sigma_3)}(x_1,x_2,x_3)$
has to be antisymmetric with respect to $x_1$ and $x_2$). On the other hand,
if $\sigma_1 = - \sigma_2$ we must have, $\Phi^{(\sigma_1, \sigma_2, \sigma_3)}(x_2,x_1,x_3)
= - \Phi^{(\sigma_2, \sigma_1, \sigma_3)}(x_1,x_2,x_3)$. Similar relations must hold in
connection with the pairs of labels $(\sigma_2, \sigma_3)$ and $(\sigma_3, \sigma_1)$.
These relations imply, in particular, that the wave function $\Phi^{(+++)}(x_1,x_2,x_3)$
(and also $\Phi^{(---)}(x_1,x_2,x_3)$) must be fully antisymmetric in the
 three coordinates $x_1, x_2, x_3$. Finally, it is clear that in order to be energy
 eigenstates the states (\ref{forabe1}-\ref{forabe2}) must involve spatial wave functions
 $\Phi^{(\sigma_1, \sigma_2, \sigma_3)}(x_1,x_2,x_3)$ that are themselves eigenfunctions
of the Hamiltonian (\ref{hami}). In particular, the three coordinate eigenfunctions associated
with (\ref{forabe1}) must be eigenfunctions of (\ref{hami}) corresponding to the same
energy eigenvalue. The ground state and few excited eigenstates of the three-electron system
that we are going to study in the present work do not correspond to the form (\ref{forabe2}).
Thus, we are going to restrict our considerations to eigenstates of the form (\ref{forabe1}).
A direct way to construct the ground and first few excited states according to the structure
(\ref{forabe1}) is to use combinations of the forms,

\[
 |n_1n_2n_3\rangle = {\cal N} \Bigl[
\Bigl( |n_{R_1}n_{R_2}n_{R_3}\rangle - |n_{R'_1}n_{R'_2}n_{R'_3}\rangle \Bigr) |++-\rangle \, +
\Bigl(|n_{R''_1}n_{R''_2}n_{R''_3}\rangle - |n_{R_1}n_{R_2}n_{R_3}\rangle\Bigr)|+-+\rangle \, +
\]

\begin{equation}
\label{spincombination1}
+ \, \Bigl(|n_{R'_1}n_{R'_2}n_{R'_3}\rangle - |n_{R''_1}n_{R''_2}n_{R''_3} \rangle \Bigr) |-++\rangle \Bigr]
\end{equation}
or
\begin{equation}
\label{spincombination2}
|n_1n_2n_3\rangle
= {\cal N}^{\prime} \,
\Bigl[
|n_{R''_1}n_{R''_2}n_{R''_3}\rangle |++-\rangle
+ |n_{R'_1}n_{R'_2}n_{R'_3}\rangle |+-+\rangle
+ |n_{R_1}n_{R_2}n_{R_3}\rangle |-++\rangle
\Bigr],
\end{equation}
where
\[
 R_1=R'_1=R''_1 , \,  R'_2=\frac{1}{\sqrt{6}}(x_1-2x_2+x_3) , \,  R''_2=\frac{1}{\sqrt{6}}(x_1+x_2-2x_3) , \,
\]
\begin{equation}
R'_3=\frac{1}{\sqrt{2}}(x_3-x_1) , \,  R''_3=\frac{1}{\sqrt{2}}(x_1-x_2),
\end{equation}
\[
 n_1=n_{R_1}=n_{R'_1}=n_{R''_1} , \, n_2=n_{R_2}=n_{R'_2}=n_{R''_2} , \, n_3=n_{R_3}=n_{R'_3}=n_{R''_3},
\]
and ${\cal N}$, ${\cal N}^{\prime}$ are appropriate normalization constants.
Note that the three spatial wave functions corresponding
respectively to the three kets $|n_{R_1}n_{R_2}n_{R_3}\rangle$,
$|n_{R'_1}n_{R'_2}n_{R'_3}\rangle$ and $|n_{R''_1}n_{R''_2}n_{R''_3}\rangle$
(which appear in (\ref{spincombination1}) and in (\ref{spincombination2}))
are obtained via cyclic permutations of the particles coordinates
$x_1,x_2,x_3$ in the definition of the Jacobi coordinates. Therefore,
it is evident that these three spatial wave functions are eigenfunctions
of (\ref{hami}) sharing the same eigenenergy. We will use combinations of type
\eqref{spincombination1} if the quantum number $n_3$ is even, and of type
\eqref{spincombination2} when it is odd, ensuring in this way the antisymmetry
of the wave function. As already mentioned, we have chosen these states
because they are also special in the sense that they all are eigenstates of $S_z$.
States of the forms (\ref{spincombination1}) and  (\ref{spincombination2})
correspond to a wave function with total spin
$S_z=S_{z}^{(1)}+S_{z}^{(2)}+S_{z}^{(3)}=+\frac{1}{2}$
of the three-electron system but one can also construct eigenstates of the
same type with total spin $S_z=-\frac{1}{2}$. As the entanglement of the $S_z=-\frac{1}{2}$
states is the same as the entanglement of states with $S_z=+\frac{1}{2}$,
in the rest of this work we will mainly focus on states with $S_z=+\frac{1}{2}$.

We must remember that these states are written in Jacobi relative coordinates
of a three-particle system and the quantum numbers $n_1$, $n_2$ and $n_3$
refer to these coordinates. However, to determine the amount of entanglement
between the particles we have to express the wave functions associated with the
eigenstates in terms of the coordinates and spins of the particles,
\begin{equation}
\label{geneigenstate}
 \Psi_{n_1n_2n_3} (x_1\sigma_1,x_2\sigma_2,x_3\sigma_3) =
\langle x_1\sigma_1,x_2\sigma_2,x_3\sigma_3 | n_1n_2n_3 \rangle
\end{equation}
In the case of the eigenfunctions \eqref{geneigenstate} of the Moshinsky system the entanglement measure $\varepsilon$  can be computed in an exact analytical way.  However, for highly excited states the corresponding expressions become very awkward. Therefore, we are going to calculate this quantity only for the ground state and the first and second excited states. In each case  we shall provide the final closed expressions for $\varepsilon$ (arising from the evaluation of the aforementioned integrals) and discuss the behavior of the eigenstate's entanglement. The value of $\varepsilon$ corresponding to the state $|n_1n_2n_3\rangle$ (with $S_z=+\frac{1}{2}$) will be denoted by $\varepsilon_{n_1n_2n_3}$. As a compact alternative notation for the alluded state we shall also use  $|n_1n_2n_3\rangle_{R_1R_2R_3}$.

We compute the state's entanglement in terms of the dimensionless parameter $\tau = \frac{\lambda}{\omega}$, which constitutes a measure of the relative strength of the interaction between two particles in the Moshinsky system. Remark that the system is decoupled when $\tau=0$. The larger the value of $\tau$, the larger is the (relative) contribution of the interaction term in the Moshinsky atom.

\subsection{Ground state $\mathbf{|010\rangle_{R_1R_2R_3}}$}

Let $A=\sqrt{1 \pm 3\tau^2}$, and using the right spin combination given by \eqref{spincombination1} we can express the entanglement of the ground state in terms of the parameter $\tau$ as
\begin{equation}
\label{tr010}
\varepsilon_{010} =1- \frac{\sqrt{2A+5A^2+2 A^3}}{4 \left(2+5 A+2 A^2\right)^3} \left(59+232 A+390 A^2+232 A^3+59 A^4\right),
\end{equation}

We see from \eqref{tr010} that the entanglement of the ground state depends upon the parameters of the Moshinsky atom only through the dimensionless quantity $\tau$. Decoupling the system, that is, making $\tau \rightarrow 0$ (which corresponds, for instance, to $\lambda \rightarrow 0$ or equivalently $\Lambda \rightarrow \omega$) makes $\varepsilon_{010}=0$ showing that in the decoupled system the ground state is not entangled. On the other hand, with maximum coupling $\tau \rightarrow \infty$ ($\tau \rightarrow \frac{1}{\sqrt{3}}$) for attractive (repulsive) interactions we find that $\varepsilon_{010}=1$, that is, the entanglement measure adopts its maximum possible value.

\subsection{First excited states $\mathbf{|110\rangle_{R_1R_2R_3}}$ and $\mathbf{|011\rangle_{R_1R_2R_3}}$}

The first excited state in energy, when the system is coupled ($\tau > 0$) and with attractive interaction, is $|110\rangle_{R_1R_2R_3}$ and the next one with higher energy is $|011\rangle_{R_1R_2R_3}$, the excitation order is reversed in the case of repulsive interaction. Both states have the same energy when we decouple the system, that is, when $\tau \rightarrow 0$. For these states, using \eqref{spincombination1} and \eqref{spincombination2} respectively, we have
\[
 \varepsilon_{110}=1-\frac{A^{1/2}}{4 \left(2+A\right)^{9/2} \left(1+2 A\right)^{9/2}} \times
\]
\begin{equation}
\times \left(177+1034 A+6213 A^2+12582 A^3+15392 A^4+12582 A^5+6213 A^6+1034 A^7+177 A^8\right)
\end{equation}
and
\[
\varepsilon_{011} = 1- \frac{A^{1/2}}{640 \left(2+A\right)^{9/2} \left(1+2 A\right)^{9/2}} \times
\]
\begin{equation}
 \times \left(3057+24608 A+93180 A^2+196704 A^3+251366 A^4+196704 A^5+93180 A^6+24608 A^7+3057 A^8\right).
\end{equation}

Decoupling the system makes $\displaystyle{\varepsilon_{011}=\varepsilon_{110}=\frac{8}{27}}$, so in the limit of a decoupled system the first excited states are entangled. On the other hand, with maximum coupling we find that $\varepsilon_{011}=\varepsilon_{110}=1$, that is, the entanglement is maximum.

To both states ($|011\rangle_{R_1R_2R_3}$ and $|110\rangle_{R_1R_2R_3}$) having $S_z=+\frac{1}{2}$, which we will denote by $|011\rangle_+$ and $|110\rangle_+$, one can associate the states $|011\rangle_-$ and $|110\rangle_-$ respectively with the same energy and same entanglement but with $S_z=-\frac{1}{2}$. Then, as these are degenerate states because the energy does not depend on the spin, we compute the amount of entanglement of a combination of them in the following way:
\[
 |\Psi_{011}\rangle \, = \, \cos \theta \, |011\rangle_+ \, + \, \sin \theta \, |011\rangle_-
\]
\begin{equation} \label{teta}
|\Psi_{110}\rangle \, = \, \cos \theta  \, |110\rangle_+ \, + \, \sin \theta \, |110\rangle_-
\end{equation}
\noindent
where $0\leq \theta < 2\pi$. These states exhibit an amount of entanglement
that is independent of the parameter $\theta$. To understand this behaviour
let us consider the unitary transformation $U$ (acting on the single-particle
Hilbert space) defined by,
\begin{eqnarray}
U |\phi_k\rangle|+\rangle \, &=& \, |\phi_k\rangle|p\rangle, \,\,\,\, k=1,2,\ldots \cr
U |\phi_k\rangle|-\rangle \, &=& \, |\phi_k\rangle|n\rangle,  \,\,\,\, k=1,2,\ldots,
\end{eqnarray}
where $(|\varepsilon_k \rangle = |\phi_k\rangle|\pm\rangle, \,\,\, k=1,2,\ldots)$
is a single-particle orthonormal basis (with the kets $|\phi_k\rangle$ corresponding
to the spatial degrees of freedom) and
\begin{eqnarray}
|p\rangle \, &=& \, \cos \theta |+\rangle - \sin \theta |-\rangle, \cr
|n\rangle \, &=& \, \sin \theta |+\rangle + \cos \theta |-\rangle.
\end{eqnarray}
It can be verified after some algebra that,
\begin{eqnarray}
|\Psi_{011}\rangle \, = \,  \Bigl(U \otimes U \otimes U \Bigr) |011\rangle_+, \cr
|\Psi_{110}\rangle  \, = \,  \Bigl(U \otimes U \otimes U \Bigr) |110\rangle_+.
\end{eqnarray}
Now, it is clear that the amount of entanglement of a three-fermions
state does not change under the effect of unitary transformations
of the form $U \otimes U \otimes U $ and, consequently, the entanglement
of the states defined in (\ref{teta}) does not depend upon $\theta$.

\subsection{Second excited states $\mathbf{|210\rangle_{R_1R_2R_3}}$, $\mathbf{|111\rangle_{R_1R_2R_3}}$, $\mathbf{|012\rangle_{R_1R_2R_3}}$, $\mathbf{|021\rangle_{R_1R_2R_3}}$ and $\mathbf{|003\rangle_{R_1R_2R_3}}$}

For these states we have that the lowest-energy second excited state when the system is coupled and with attractive interaction is $|210\rangle_{R_1R_2R_3}$, the next one with higher energy is $|111\rangle_{R_1R_2R_3}$, and the following three states, all of them with the same energy, are $|012\rangle_{R_1R_2R_3}$, $|021\rangle_{R_1R_2R_3}$ and $|003\rangle_{R_1R_2R_3}$. All these states have the same energy when the system is decoupled.

Defining the parameter $B=\frac{A^{1/2}}{\left(2+A\right)^{13/2} \left(1+2 A\right)^{13/2}}$, using Eqs. \eqref{spincombination1} for the states $|210\rangle_{R_1R_2R_3}$ and $|012\rangle_{R_1R_2R_3}$, and \eqref{spincombination2} for the states $|111\rangle_{R_1R_2R_3}$, $|021\rangle_{R_1R_2R_3}$ and $|003\rangle_{R_1R_2R_3}$, we found that
\[
 \varepsilon_{210} = 1- \frac{B}{16} \left(2419+19480 A+218138 A^2+564200 A^3+1466241 A^4+2943840 A^5+3743124 A^6+\right.
\]
\[
\left. +2943840 A^7+1466241 A^8+564200 A^9+218138 A^{10}+19480 A^{11}+2419 A^{12}\right),
\]
\[
 \varepsilon_{111} = 1-\frac{B}{64} \left(9171+80546 A+700555 A^2+ 2659770 A^3+6668841 A^4+11416740 A^5+13615794 A^6+\right.
\]
\[
\left. +11416740 A^7+6668841 A^8+2659770 A^9+700555 A^{10}+80546 A^{11}+9171 A^{12}\right),
\]
\[
 \varepsilon_{012} = 1- \frac{B}{256} \left(42739+506008 A+3123242 A^2+11179160 A^3+26922957 A^4+44982480 A^5+53234988 A^6+\right.
\]
\[
\left. +44982480 A^7+26922957 A^8+11179160 A^9+3123242 A^{10}+506008 A^{11}+42739 A^{12}\right),
\]
\[
 \varepsilon_{021} = 1-\frac{B}{4096} \left(727363+8982520 A+54219206 A^2+196856600 A^3+469858317 A^4+776694000 A^5+\right.
\]
\[
\left. +915625428 A^6+776694000 A^7+469858317 A^8+196856600 A^9+54219206 A^{10}+8982520 A^{11}+727363 A^{12}\right),
\]
and
\[
 \varepsilon_{003} = 1-\frac{B}{4096} \left(762395+9419160 A+61156086 A^2+232139320 A^3+576896949 A^4+982782000 A^5+\right).
\]
\[
\left. +1171448436 A^6+982782000 A^7+576896949 A^8+232139320 A^9+61156086 A^{10}+9419160 A^{11}+762395 A^{12}\right),
\]

Taking the limit for the decoupling case of the system, makes $\displaystyle{\varepsilon_{111}=\varepsilon_{210}=\varepsilon_{012}=\frac{4}{9}}$, $\displaystyle{\varepsilon_{021}=\frac{43}{108}}$ and $\displaystyle{\varepsilon_{003}=\frac{1}{4}}$ showing again that these excited states are entangled in the decoupled system. In the maximum coupling limit we find for all second-excited states that the entanglement reaches again its maximum value.

\begin{figure}[h]
\begin{center}
\includegraphics[scale=0.9,angle=0]{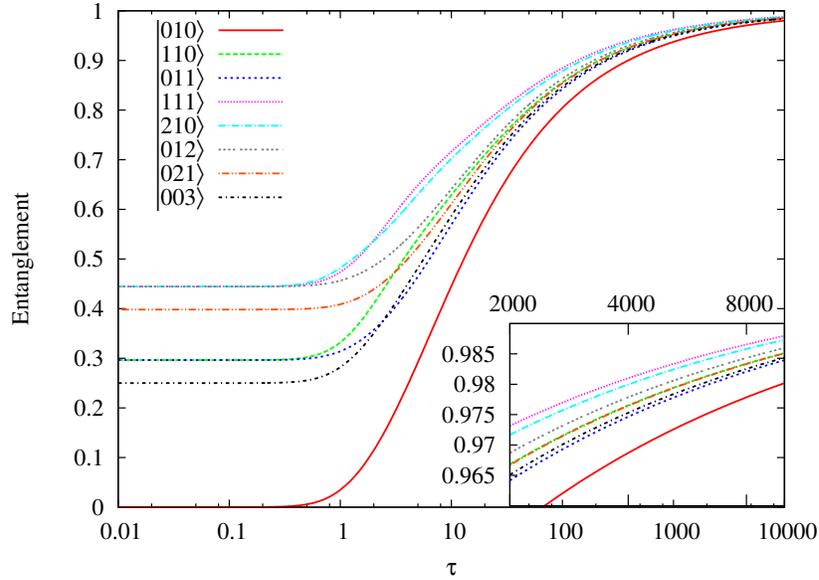}
\caption{(Color online) Entanglement of the ground, first and second excited states of one-dimensional Moshinsky atom with three electrons attractively interacting. All depicted quantities are dimensionless.\label{figure_1}}
\end{center}
\end{figure}

\begin{figure}[h]
\begin{center}
\includegraphics[scale=0.9,angle=0]{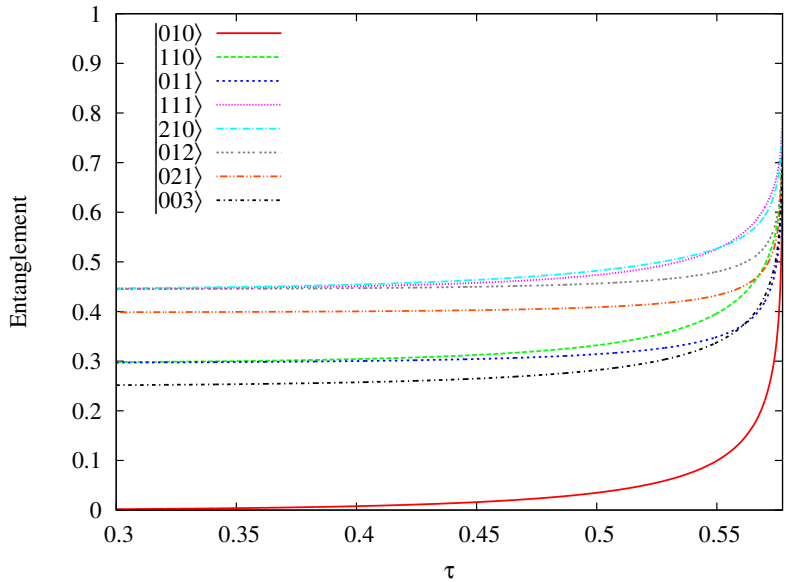}
\caption{(Color online) Entanglement of the ground, first and second excited
states of one-dimensional Moshinsky atom with three interacting electrons for
the repulsive case. All depicted quantities are dimensionless. \label{figure_2}}
\end{center}
\end{figure}

The behaviour of the eigenstates' entanglement as a function of the parameter $\tau$ (which corresponds to the relative strength of the
interaction between the two particles) is depicted in Figure 1 for an attractive interaction and in Figure 2 for a repulsive interaction.

Comparing Figures 1 and 2 one observes that in the repulsive case
(Fig.2) maximum entanglement is reached when the parameter $\tau$
approaches the finite limit value $\frac{1}{\sqrt{3}} \approx 0.577$.
In the attractive case (Fig.1) entanglement behaves in a different way:
maximum entanglement corresponds to the limit $\tau \to \infty$. This
difference between the attractive and the repulsive cases is due to the
fact that the Moshinsky model with repulsive interaction admits bound
states only for $\tau$-values in the finite range $[0,\frac{1}{\sqrt{3}})$.
On the other hand, in the attractive case the Moshinsky model admits
bound states for all $\tau \ge 0$. In the case of the repulsive
interaction the eigenstates of the system are no longer bounded
for $\tau \ge \frac{1}{\sqrt{3}}$. Thus, the eigenstates exhibit
a qualitative structural change at the ``critical'' value
$\tau_c = \frac{1}{\sqrt{3}}$, resembling a quantum phase
transition. A similar situation occurs in the case of the
Moshinsky atom with two electrons in a uniform magnetic field
studied in the next section (see Figure 4). This system,
when the interaction is repulsive, admits bound states
only for $\tau$-values smaller than the critical value $\tau_c=1$.

\section{Two-electron Moshinsky atom in a magnetic field}

The Hamiltonian of the two electron Moshinsky atom in a three dimensional space is
\[
 H_M = \frac{p_1^2}{2m_e}+\frac{p_2^2}{2m_e}+ \frac{1}{2}m_e\omega^2(r_1^2+r_2^2) \pm \frac{\lambda^2}{2}(\mathbf{r}_1-\mathbf{r}_2)^2
\]
where subscripts $1$ and $2$ denote each of the electrons. As before, the positive (negative) sign refers to a attractive (repulsive) interaction between the electrons. To study the presence of a uniform magnetic field acting on the system, we perform the following change in the Hamiltonian
\[
 \mathbf{p}_1 \rightarrow \mathbf{p}_1+\frac{e}{c}\mathbf{A} \text{ \ \ and \ \ }  \mathbf{p}_2 \rightarrow \mathbf{p}_2+\frac{e}{c}\mathbf{A} \text{ \ \ with \ \ }  \mathbf{A} = \frac{1}{2} (\mathbf{B}\wedge \mathbf{r})
\]
being $\mathbf{B}$ the magnetic field. Assuming that the magnetic field is homogeneous and have $z$-axis direction, that is $\mathbf{B}=B\hat{z}$, we can write:
\begin{equation}
\label{pmagnetico}
 p_i^2 \rightarrow p_i^2 + \left(\frac{eB}{2c}\right)^2(x_i^2+y_i^2) + \frac{eB}{c}(x_ip_{yi}-y_ip_{xi}) \text{ \ \ with \ \ } i=1,2
\end{equation}

By replacing \eqref{pmagnetico} in the Hamiltonian $H_M$ and setting atomic units ($m_e=\hbar=1$, $c=1/\alpha$), we obtain
\begin{equation}
\label{hamiltoniano1}
 H = \frac{1}{2} (p_1^2+p_2^2) + \frac{\omega^2}{2} (r_1^2+r_2^2) + \frac{b^2}{2} (x_1^2+y_1^2+x_2^2+y_2^2)+ b (L_{1z}+L_{2z}) \pm \frac{\lambda}{2} (\mathbf{r}_1-\mathbf{r}_2)^2
\end{equation}
where
\begin{equation}
\label{parametros}
 b=\frac{B}{2c} \text{ \ \ ; \ \ } L_{iz}=(x_ip_{yi}-y_ip_{xi}) \text{ \ \ and \ \ } \mathbf{r}_i = (x_i,y_i,z_i) \text{ \ \ with \ \ } i=1,2
\end{equation}

We change the variables to the center of mass (CM) and relative coordinates, i.e.
\begin{equation}
\label{cambiocoordenadas}
 \mathbf{R} = \frac{1}{\sqrt{2}} (\mathbf{r}_1+\mathbf{r}_2) \text{ \ \ y \ \ } \mathbf{r} = \frac{1}{\sqrt{2}} (\mathbf{r}_1-\mathbf{r}_2)
\end{equation}
respectively. This transformation satisfies the relations
\begin{equation}
\label{relaciontransformaciones}
 p_1^2+p^2_2 = p_R^2+p_r^2 \text{ \ \ and \ \ } L_{1z}+L_{2z} = L_{Rz}+L_{rz} = (R_xp_{Ry}-R_yp_{Rx})+(r_xp_{ry}-r_yp_{rx}),
\end{equation}
and therefore, introducing equations \eqref{cambiocoordenadas} and \eqref{relaciontransformaciones} in the Hamiltonian \eqref{hamiltoniano1} we obtain
\begin{equation}
 \label{hamiltoniano2}
H =  \frac{1}{2} (p_R^2+p_r^2) + \frac{\omega^2}{2} (R^2+r^2) + \frac{b^2}{2} (R_x^2+R_y^2+r_x^2+r_y^2) + b (L_{Rz}+L_{rz}) \pm \frac{\lambda^2}{2} r^2,
\end{equation}
which is separable in the CM and relative coordinates, so that we can express as
\[
 H = H_R + H_r
\]
where
\begin{equation}
 \label{HR1}
H_R=\frac{1}{2} (p_{Rx}^2+p_{Ry}^2) + \frac{\omega^2+b^2}{2} (R_x^2+R_y^2) + \frac{1}{2} (p_{Rz}^2 + \omega^2 R_z^2) + b L_{Rz}
\end{equation}
and
\begin{equation}
 \label{Hr1}
H_r=\frac{1}{2} (p_{rx}^2+p_{ry}^2) + \frac{\omega^2+b^2}{2} (r_x^2+r_y^2) + \frac{1}{2} (p_{rz}^2 + \omega^2 r_z^2) + b L_{rz} \pm \frac{\lambda}{2} r^2.
\end{equation}

Introducing the following dilation canonical transformation for the Hamiltonian $H_R$
\[
 p'_{Ri} = (\omega^2+b^2)^{-\frac{1}{4}} p_{Ri} , \, R'_i = (\omega^2+b^2)^{\frac{1}{4}} R_i \text{ \ \ with \ \ } i = x, y
\]
\begin{equation}
\label{jacobian1}
 p'_{Rz} = \omega^{-\frac{1}{2}} p_{Rz} , \, R'_z = \omega^{\frac{1}{2}} R_z,
\end{equation}
and
\[
 p'_{ri} = (\omega^2+b^2\pm\lambda^2)^{-\frac{1}{4}} p_{ri} , \, r'_i = (\omega^2+b^2\pm\lambda^2)^{\frac{1}{4}} r_i \text{ \ \ with \ \ } i = x, y
\]
\begin{equation}
\label{jacobian2}
 p'_{rz} = (\omega^2\pm\lambda^2)^{-\frac{1}{4}} p_{rz} , \, r'_z = (\omega^2\pm\lambda^2)^{\frac{1}{4}} r_z,
\end{equation}
we obtain
\begin{equation}
 \label{HR2}
H'_R = \frac{H_R}{\omega} = \frac{1}{2} \left(1+\frac{b^2}{\omega^2}\right)^{\frac{1}{2}} ({p'_{Rx}}^2+{p'_{Ry}}^2+{R'_x}^2+{R'_y}^2)+\frac{1}{2}({p'_{Rz}}^2+{R'_z}^2) + \frac{b}{\omega} L_{R'z}
\end{equation}
\begin{equation}
 \label{Hr2}
H'_r = \frac{H_r}{(\omega^2\pm\lambda^2)^\frac{1}{2}} = \frac{1}{2} \left(1+\frac{b^2}{\omega^2\pm\lambda^2}\right)^{\frac{1}{2}} ({p'_{rx}}^2+{p'_{ry}}^2+{r'_x}^2+{r'_y}^2)+\frac{1}{2}({p'_{rz}}^2+{r'_z}^2) + \frac{b}{(\omega^2\pm\lambda^2)^\frac{1}{2}} L_{r'z}
\end{equation}

The Hamiltonian describing the whole system will be therefore
\begin{equation}
\label{Hamiltonianomagnetico}
 H = \omega H'_R + (\omega^2\pm\lambda^2)^\frac{1}{2} H'_r.
\end{equation}
Using cylindrical coordinates, that is
\begin{equation}
\label{cilindricasR}
 \rho_R = ({R'_x}^2+{R'_y}^2)^\frac{1}{2} , \, \varphi = \arctan \left(\frac{{R'_y}^2}{{R'_x}^2}\right) , \, z_R=R'_z,
\end{equation}
\begin{equation}
\label{cilindricasr}
 \rho_r = ({r'_x}^2+{r'_y}^2)^\frac{1}{2} , \, \varphi = \arctan\left(\frac{{r'_y}^2}{{r'_x}^2}\right) , \, z_r=r'_z,
\end{equation}
we immediately have the eigenfunctions of $H'_R$ y $H'_r$ given by \cite{MMM}
\begin{equation}
 \Psi_{\nu_R m_R n_R} (\mathbf{R}) = \frac{1}{\sqrt{2\pi}} R_{\nu_R|m_R|}(\rho_R) e^{im_R\varphi_R} \chi_{n_R}(z_R)
\end{equation}
\begin{equation}
 \Psi_{\nu_R m_R n_R} (\mathbf{r}) = \frac{1}{\sqrt{2\pi}} R_{\nu_r|m_r|}(\rho_r) e^{im_r\varphi_r} \chi_{n_r}(z_r),
\end{equation}
where $R_{\nu|m|}(\rho)$ are the two-dimensional oscillator radial eigenstates, whose normalized expressions are
\begin{equation}
 R_{\nu|m|}(\rho) = \left(\frac{2\ \nu!}{(\nu+|m|)!} \right)^\frac{1}{2} \rho^{|m|} e^{-\frac{\rho}{2}} L_\nu^{|m|}(\rho^2)
\end{equation}
being  $L_\nu^{|m|}$ the Laguerre polynomials with the quantum numbers $\nu$ and $m$ taking the values $\nu=0,1,2,...$ and $m=0,\pm1,\pm2,...$ respectively. The functions $\chi_\tau(z)$ are the eigenstates of the unidimensional harmonic oscillator which are given by
\begin{equation}
 \chi_n(z) = \left(\frac{1}{2^n n! \pi^\frac{1}{2}}\right)^\frac{1}{2} e^{-\frac{z^2}{2}}H_n(z)
\end{equation}
where $H_n(z)$ are the Hermite polynomials and $n$ takes the values $n=0,1,2,...$

The final eigenstates of the Hamiltonian \eqref{Hamiltonianomagnetico} will be
\begin{equation}
\label{autofunciontotal}
 |\nu_R m_R n_R,\nu_r m_r n_r \rangle = |\nu_R m_R n_R \rangle \otimes |\nu_r m_r n_r \rangle
\end{equation}
and the wave function
\begin{equation}
 \Psi_{\nu_R m_R n_R,\nu_r m_r n_r} (\mathbf{r_1},\mathbf{r_2})= \Psi_{\nu_R m_R n_R,\nu_r m_r n_r} (\mathbf{R},\mathbf{r}) |J| = \langle x_1,y_1,z_1;x_2,y_2,z_2 | \nu_R m_R n_R,\nu_r m_r n_r \rangle
\end{equation}
where $J$ is the Jacobian of the canonical transformation \eqref{jacobian1} and \eqref{jacobian2}.

The eigenvalues of the harmonic oscillators in one and two dimensions are $(n+\frac{1}{2})$ and $(2\nu+|m|+1)$, respectively. Defining the quantities
\begin{equation}
 y_R = \left(1+\frac{b^2}{\omega^2}\right)^\frac{1}{2} + \frac{b}{\omega} \text{ \ \ \ \ and \ \ \ \ } y_r = \left(1+\frac{b^2}{\omega^2\pm\lambda^2}\right)^\frac{1}{2} + \frac{b}{(\omega^2\pm\lambda^2)^\frac{1}{2}},
\end{equation}
we obtain the eigenvalues of the Hamiltonians $H'_R$ and $H'_r$ in the form
\begin{equation}
 E'_{\nu_Rm_Rn_R}(y_R) = \frac{y_R}{2} (2\nu_R+|m_R|+m_R+1) + \frac{1}{2y_R} (2\nu_R+|m_R|-m_R+1) + \left(n_R+\frac{1}{2}\right)
\end{equation}
\begin{equation}
 E'_{\nu_rm_rn_r}(y_r) = \frac{y_r}{2} (2\nu_r+|m_r|+m_r+1) + \frac{1}{2y_r} (2\nu_r+|m_r|-m_r+1) + \left(n_r+\frac{1}{2}\right)
\end{equation}
Then, the total energy of the system, which is the eigenvalue of the Hamiltonian $H$, is given by
\begin{equation}
\label{totalenergy}
E_{\nu_Rm_Rn_R\nu_rm_rn_r}(\omega,b) = \omega E'_{\nu_Rm_Rn_R}(y_R) + (\omega^2\pm\lambda^2)^\frac{1}{2} E'_{\nu_rm_rn_r}(y_r)
\end{equation}

We calculate the exact form of the trace of the reduced density matrix associated to a general eigenfunction \eqref{autofunciontotal} of the two-electron Moshinsky system with magnetic field for the ground and the first excited states in $n_R$, $n_r$, $\nu_R$ and $\nu_r$. Next we are going to provide and discuss the corresponding amounts of entanglement exhibited by each eigenstate (arising from the evaluation of the aforementioned integrals). In what follows, $\varepsilon_{\nu_R m_R n_R,\nu_r m_r n_r}$ denotes value of $\varepsilon$ when evaluated on the state $|\nu_R m_R n_R,\nu_r m_r n_r \rangle$ that we also will denote $|\nu_R m_R n_R, \nu_r m_r n_r \rangle_{Rr}$. In order to obtain physically acceptable solutions in the case of a repulsive interaction between the particles we have to take into account the constraint $\lambda < \omega$.

\subsection{Ground state $\mathbf{|000,000\rangle_{Rr}}$}

The ground state is symmetric in coordinates, so we must combine it with the only antisymmetric spin function to ensure the antisymmetry of the wave function. Let $\displaystyle{\sigma = \frac{b}{\omega}}$ and $\displaystyle{\tau = \frac{\lambda}{\omega}}$ as before. In this case we have
\begin{equation}
\varepsilon_{000,000} = 1- \frac{8 \sqrt{1+\sigma ^2} \sqrt{1+\tau ^2} \sqrt{1+\sigma ^2+\tau ^2} \left(2+2 \sigma ^2+\tau ^2-2 \sqrt{1+\sigma ^2} \sqrt{1+\sigma ^2+\tau ^2}\right)}{\tau ^4 \left(1+\sqrt{1+\tau ^2}\right) \sqrt{\frac{1}{1-\frac{4}{\tau ^2}+\frac{4+5 \tau ^2}{\tau ^2 \sqrt{1+\tau ^2}}}} \sqrt{\frac{\tau ^2+2 \left(1+3 \sqrt{1+\tau ^2}\right)}{1+\sqrt{1+\tau ^2}}}}
\end{equation}

Decoupling the system makes $\displaystyle{\varepsilon_{000,000}=0}$, therefore in the decoupled system the ground state is not entangled. With maximum coupling $\tau \rightarrow \infty$ ($\tau\rightarrow 1$) in the attractive (repulsive) case, we find that $\varepsilon_{000,000}=1$; that is, the entanglement measure is maximum.
The behaviour of entanglement as a function of the parameters $\tau $ and $\sigma $ is shown in Figure 3. Figures 3a and 3b correspond,
respectively, to the attractive and the repulsive cases. More detailed information concerning the asymptotic behaviour of entanglement
is provided in Figure 4.

\begin{figure}[h]
\begin{center}
\includegraphics[scale=0.65,angle=0]{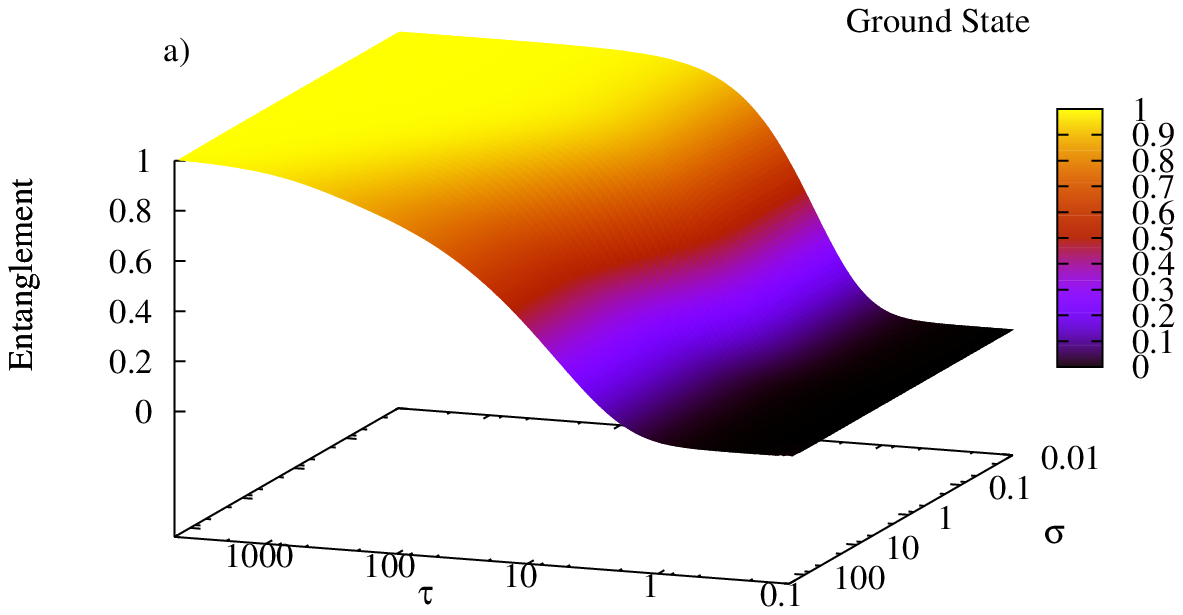}
\includegraphics[scale=0.65,angle=0]{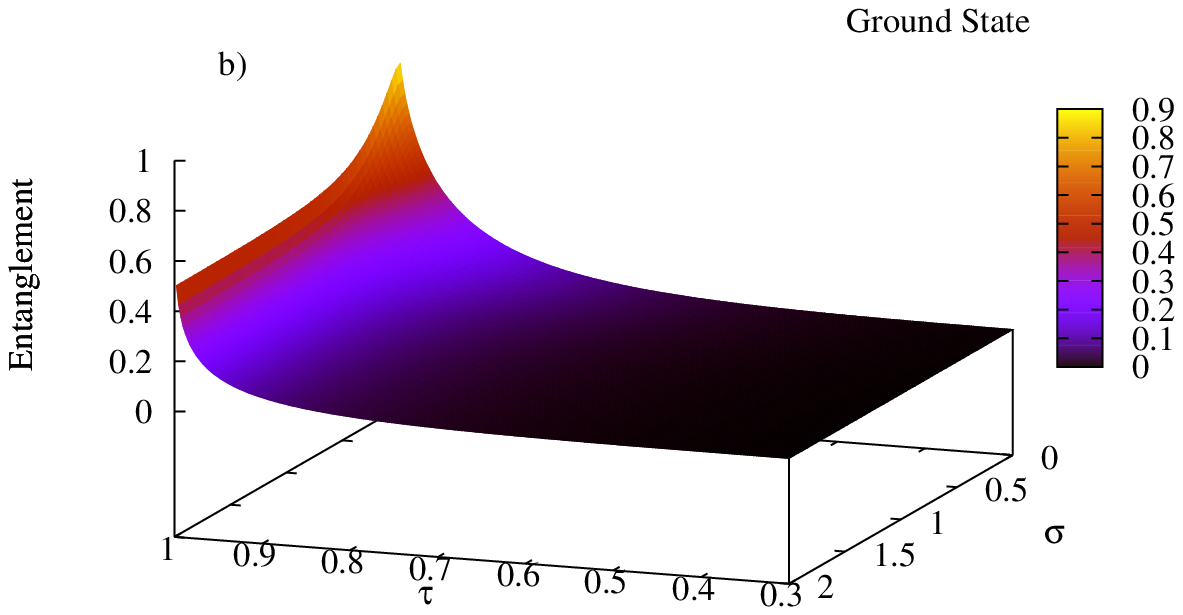}
\caption{(Color online) Entanglement of the ground state of the three-dimensional Moshinsky atom with two interacting electrons and a magnetic field. a) Attractive interaction, b) Repulsive interaction. All depicted quantities are dimensionless. \label{mag-gs}}
\end{center}
\end{figure}

From Fig. (\ref{mag-gs}) it can be observed that in the limit of large magnetic fields, that is $\sigma \rightarrow \infty$, the entanglement reaches a constant value which depends on the relative strength of interaction given by the parameter $\tau$, i.e.
\begin{equation}
 \lim_{\sigma \rightarrow \infty} \varepsilon_{000,000} = 1- \frac{8 \left(1+\tau ^2\right) \left(2+\tau ^2-2 \sqrt{1+\tau ^2}\right)}{\tau ^4 \left(1+\sqrt{1+\tau ^2}\right) \sqrt{\frac{2+\tau ^2+6 \sqrt{1+\tau ^2}}{1+\sqrt{1+\tau ^2}}} \sqrt{\frac{1}{1-\frac{4}{\tau ^2}+\frac{4+5 \tau ^2}{\tau ^2 \sqrt{1+\tau ^2}}}}}
\end{equation}

\begin{figure}[h]
\begin{center}
\includegraphics[scale=0.7,angle=0]{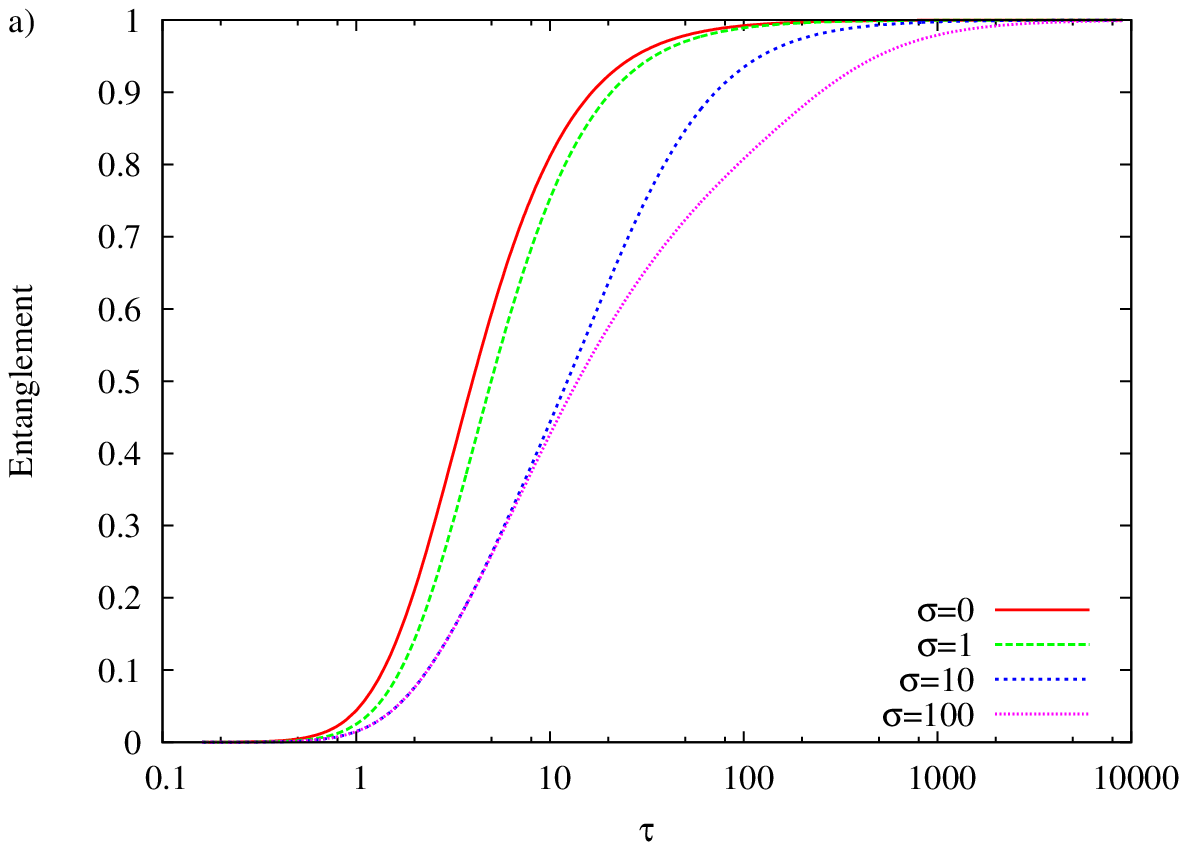}
\includegraphics[scale=0.7,angle=0]{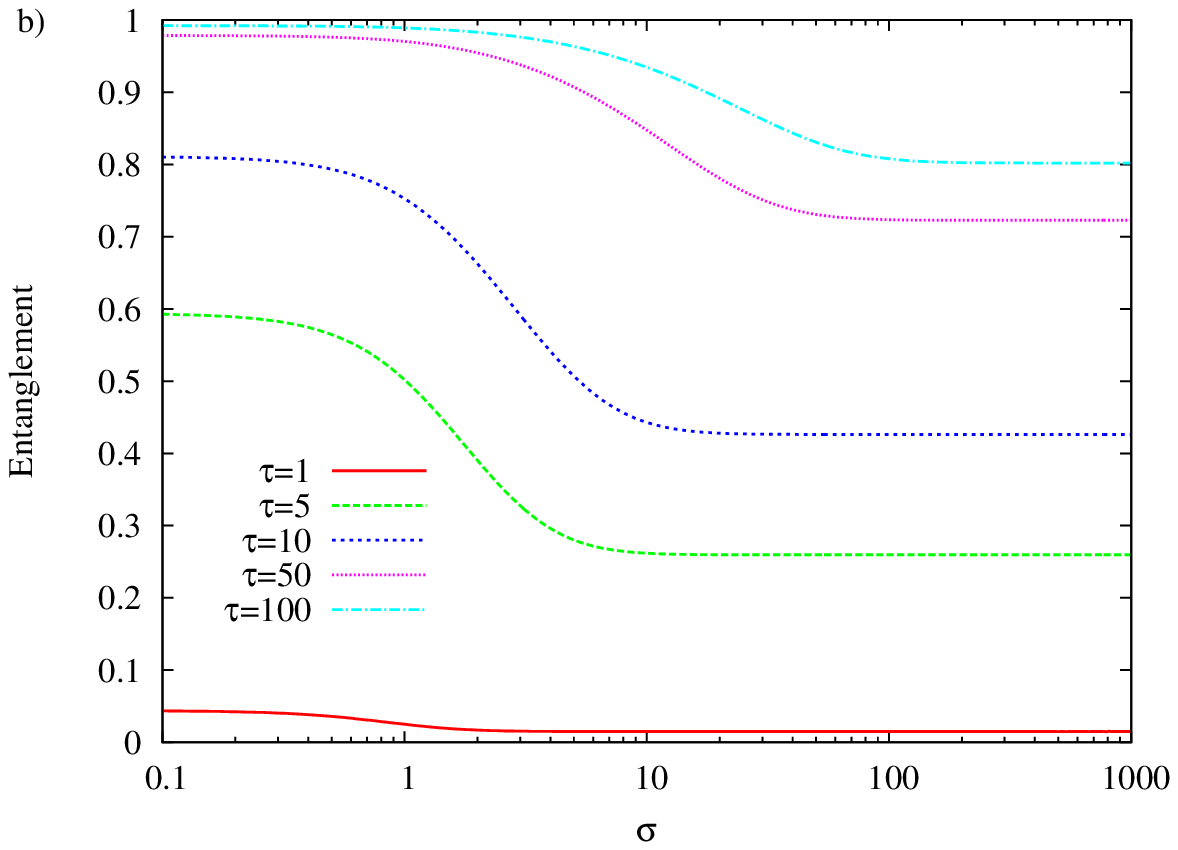}
\includegraphics[scale=0.7,angle=0]{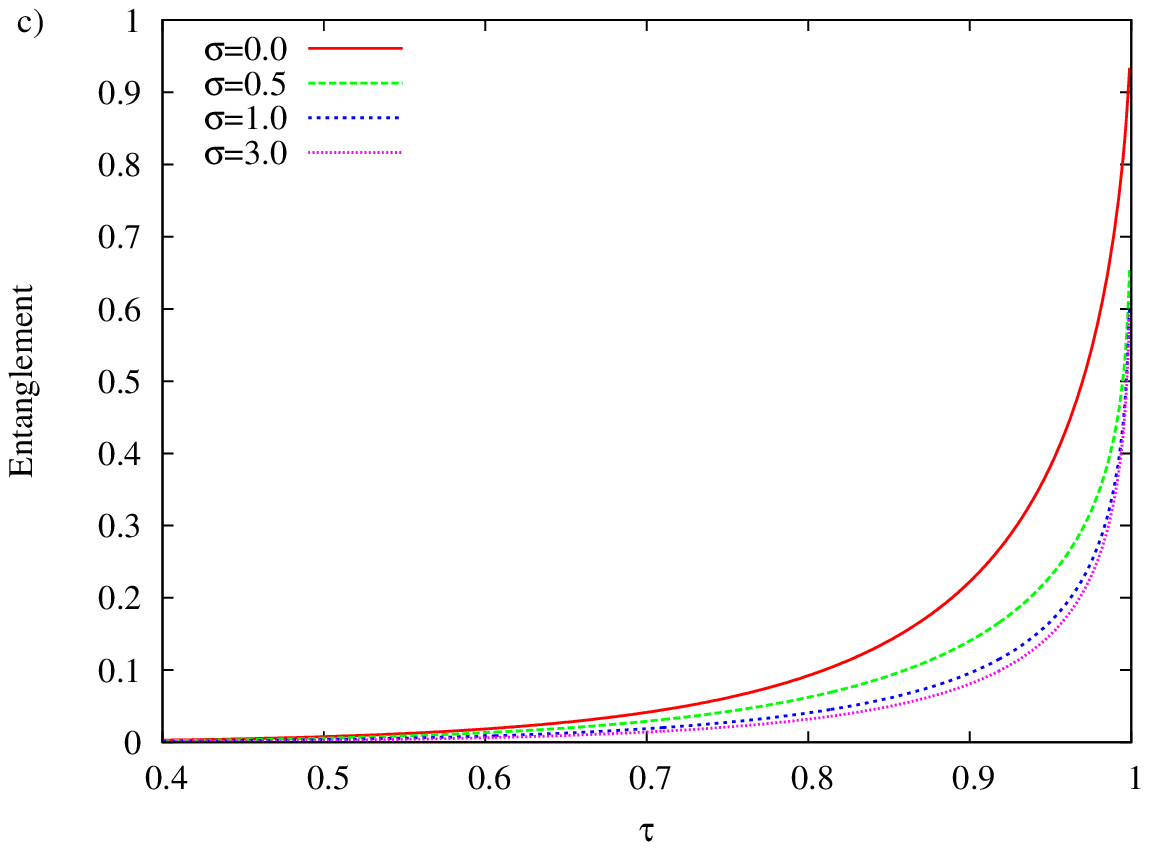}
\includegraphics[scale=0.7,angle=0]{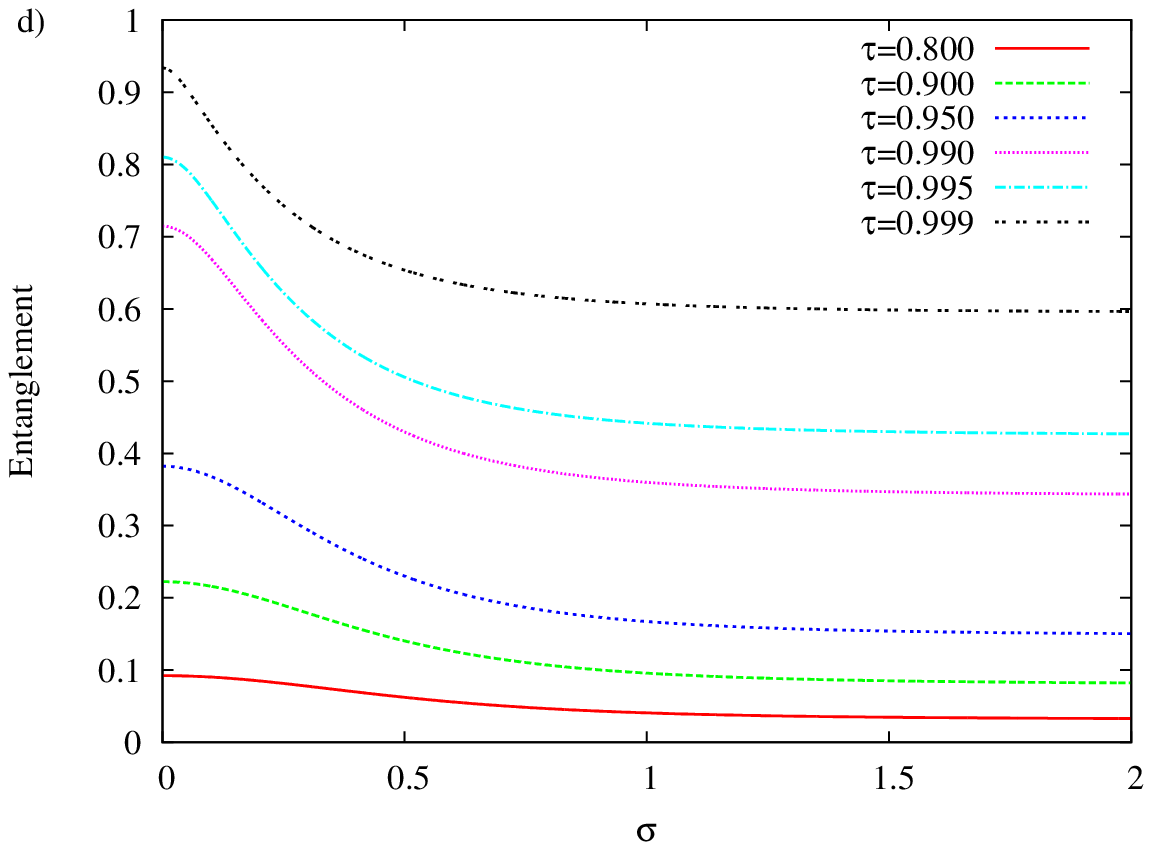}
\caption{(Color online) Entanglement of the ground state of the three-dimensional Moshinsky atom with two interacting electrons in a uniform magnetic field, a) as a function of $\tau$ for different values of $\sigma$ with attractive interaction, b) as function of $\sigma$ for different values of $\tau$ with attractive interaction, c) as a function of $\tau$ for different values of $\sigma$ with repulsive interaction and d) as a function of $\sigma$ for different values of $\tau$ with repulsive interaction. All depicted quantities are dimensionless. \label{figure_4}}
\end{center}
\end{figure}

\subsection{First excited states $\mathbf{|100,000\rangle_{Rr}}$,$\mathbf{|000,100\rangle_{Rr}}$,$\mathbf{|001,000\rangle_{Rr}}$,$\mathbf{|000,001\rangle_{Rr}}$}

We study the excited states in $\nu_R$, $\nu_r$ and $n_R$ that have symmetric coordinates wave functions and therefore must be combined with the antisymmetric or antiparallel spin function. We also study in this section the excited state in $n_r$ which is antisymmetric in coordinates and therefore, it can be combined with parallel or antiparallel spin functions. Excited eigenstates in $\nu_R$ and $\nu_r$ and in $n_R$ and $n_r$ have the same energy respectively when the system is decoupled. In this case we obtain
\begin{equation}
 \varepsilon_{000,100}=\varepsilon_{100,000} = 1- \frac{4 \sqrt{1+\sigma ^2} \sqrt{1+\tau ^2} \sqrt{1+\sigma ^2+\tau ^2} \left(8+8 \sigma ^4+8 \tau ^2+\tau ^4+8 \sigma ^2 \left(2+\tau ^2\right)\right)}{\left(1+\sqrt{1+\tau ^2}\right) \sqrt{\frac{1+\tau ^2+\sqrt{1+\tau ^2}}{2+\tau ^2+6 \sqrt{1+\tau ^2}}} \sqrt{\frac{2+\tau ^2+6 \sqrt{1+\tau ^2}}{1+\sqrt{1+\tau ^2}}} \left(\sqrt{1+\sigma ^2}+\sqrt{1+\sigma ^2+\tau ^2}\right)^6},
\end{equation}
 \begin{equation}
 \varepsilon_{000,001}=\varepsilon_{001,000} = 1- \alpha \frac{2 \sqrt{1+\sigma ^2} \sqrt{1+\sigma ^2+\tau ^2} \left(6+3 \tau ^2+2 \sqrt{1+\tau ^2}\right) \sqrt{\frac{1+\tau ^2+\sqrt{1+\tau ^2}}{2+\tau ^2+6 \sqrt{1+\tau ^2}}} \sqrt{\frac{2+\tau ^2+6 \sqrt{1+\tau ^2}}{1+\sqrt{1+\tau ^2}}}}{\left(1+\sqrt{1+\tau ^2}\right)^3 \left(\sqrt{1+\sigma ^2}+\sqrt{1+\sigma ^2+\tau ^2}\right)^2}
\end{equation}
where $\alpha=1(2)$ for antiparallel (parallel) spin.

Taking the decoupled limit system, we obtain the following entanglement values regardless of the magnetic field value: $\displaystyle{\varepsilon^a_{100,000}=\varepsilon^a_{000,100}=\frac{3}{4}}$ and $\displaystyle{\varepsilon^a_{001,000}=\varepsilon^a_{000,001}=\frac{1}{2}}$ for the first excited states with antiparallel spin, which are entangled, and $\varepsilon^p_{000,001}=0$ for the only possible state with parallel spin. We have used $\varepsilon^a$ ($\varepsilon^p$) to indicate the entanglement of states with antiparallel (parallel) spin.

On the other hand, with maximum coupling we find that $\varepsilon^a_{100,000}=\varepsilon^a_{000,100}=\varepsilon^a_{001,000}=\varepsilon^a_{000,001}=\varepsilon^p_{000,001}=1$, the entanglement is maximum.

In the limit for large magnetic fields, i.e. $\sigma \rightarrow \infty$, we obtain
\begin{equation}
 \lim_{\sigma \rightarrow \infty} \varepsilon_{100,000} = \lim_{\sigma \rightarrow \infty} \varepsilon_{000,100} = 1- \frac{3\sqrt{1+\tau ^2}}{2 \left(1+\sqrt{1+\tau ^2}\right) \sqrt{\frac{1+\tau ^2+\sqrt{1+\tau ^2}}{2+\tau ^2+6 \sqrt{1+\tau ^2}}} \sqrt{\frac{2+\tau ^2+6 \sqrt{1+\tau ^2}}{1+\sqrt{1+\tau ^2}}}}
\end{equation}
\begin{equation}
 \lim_{\sigma \rightarrow \infty} \varepsilon_{001,000} = \lim_{\sigma \rightarrow \infty} \varepsilon_{000,001} = 1-\alpha \frac{3\left(6+3 \tau ^2+2 \sqrt{1+\tau ^2}\right) \sqrt{\frac{1+\tau ^2+\sqrt{1+\tau ^2}}{2+\tau ^2+6 \sqrt{1+\tau ^2}}} \sqrt{\frac{2+\tau ^2+6 \sqrt{1+\tau ^2}}{1+\sqrt{1+\tau ^2}}}}{ \left(1+\sqrt{1+\tau ^2}\right)^3}
\end{equation}
which, as we observe, depends on the value of the interaction.
\begin{figure}[h]
\begin{center}
\includegraphics[scale=0.6,angle=0]{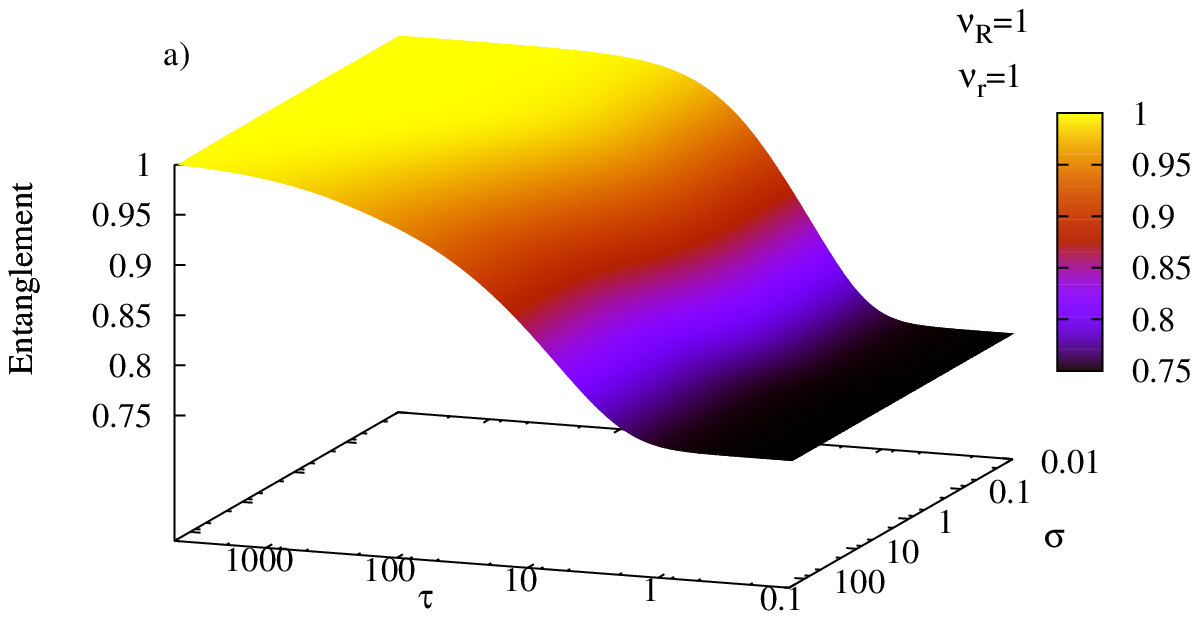}
\includegraphics[scale=0.6,angle=0]{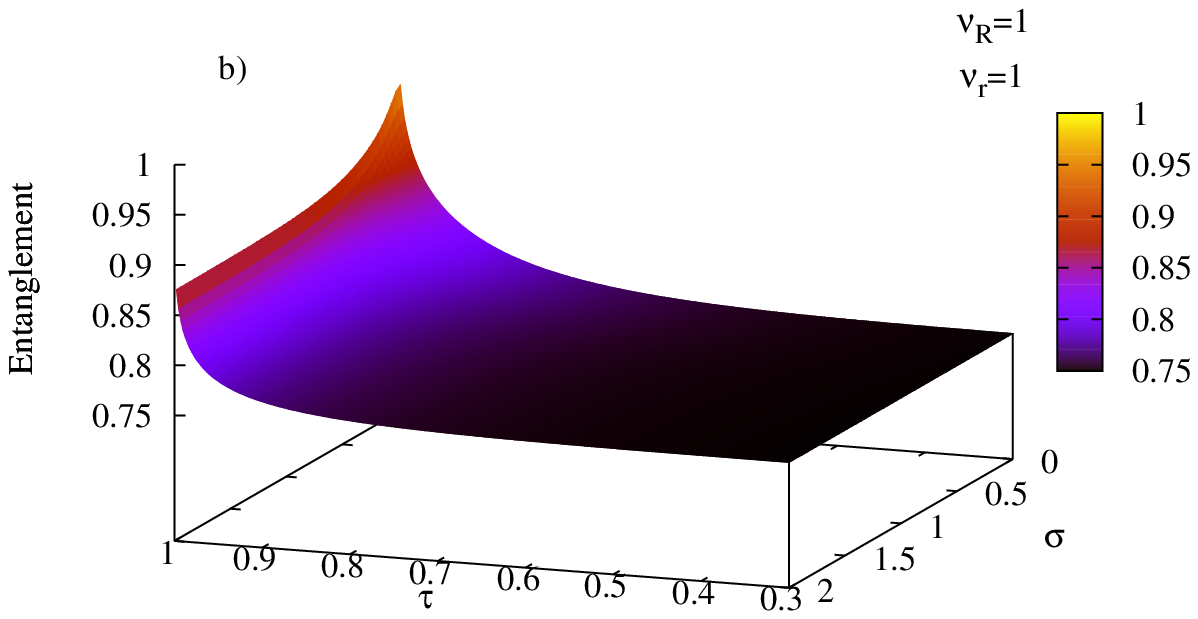}
\caption{(Color online) Entanglement of the first excited state in $\nu_R$ and $\nu_r$ of the three dimensional Moshinsky atom with two interacting electrons and magnetic field. a) Attractive interaction, b) Repulsive interaction. All depicted quantities are dimensionless. \label{figure_5}}
\end{center}
\end{figure}
\begin{figure}[h]
\begin{center}
\includegraphics[scale=0.6,angle=0]{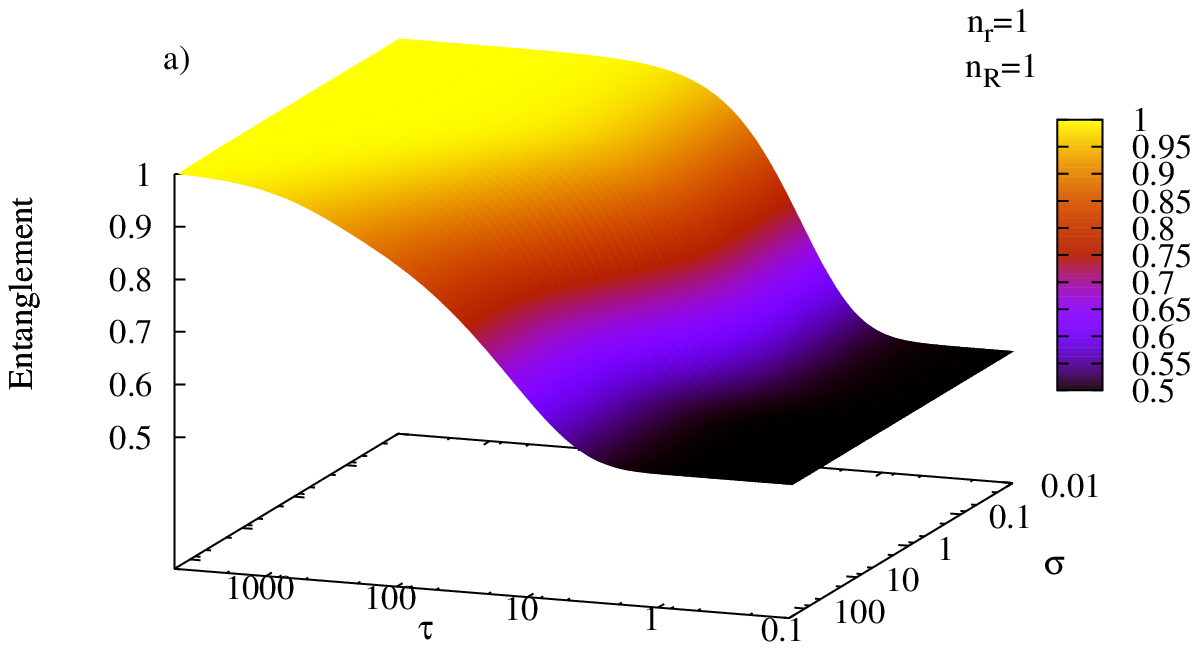}
\includegraphics[scale=0.6,angle=0]{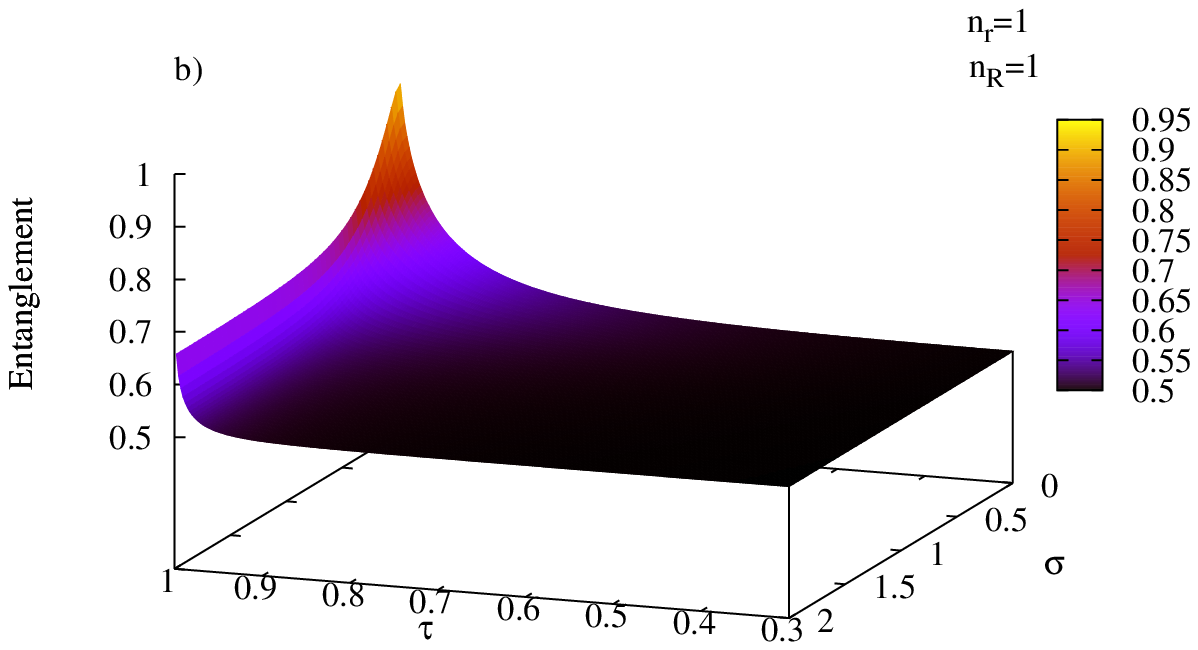}
\includegraphics[scale=0.6,angle=0]{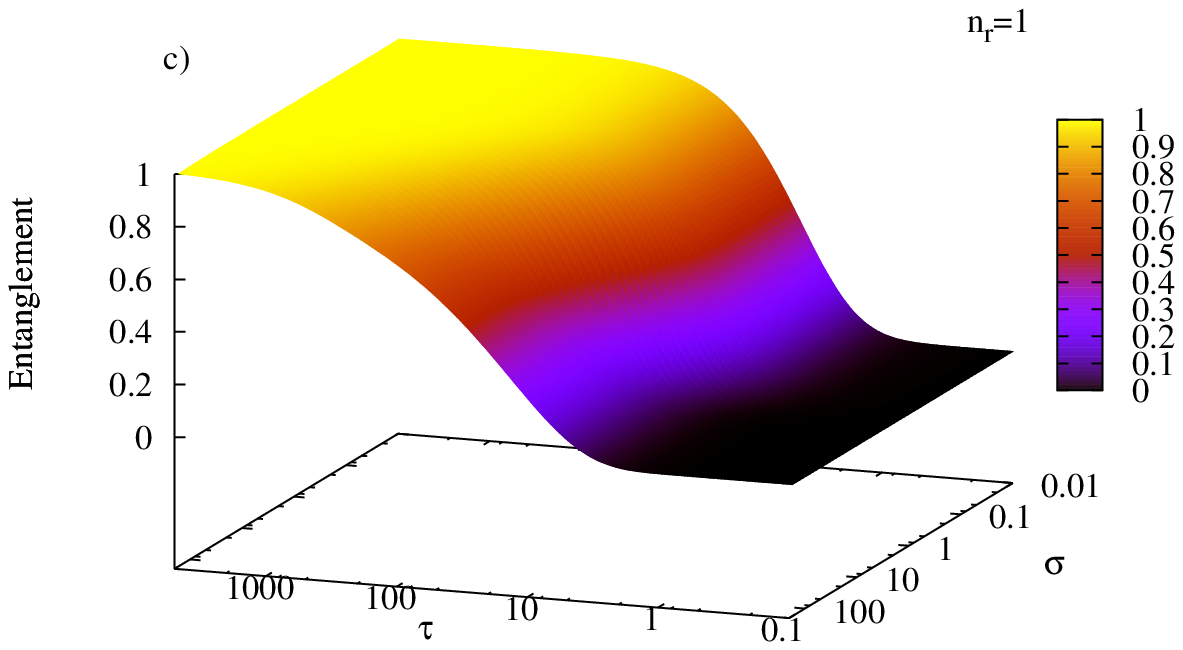}
\includegraphics[scale=0.6,angle=0]{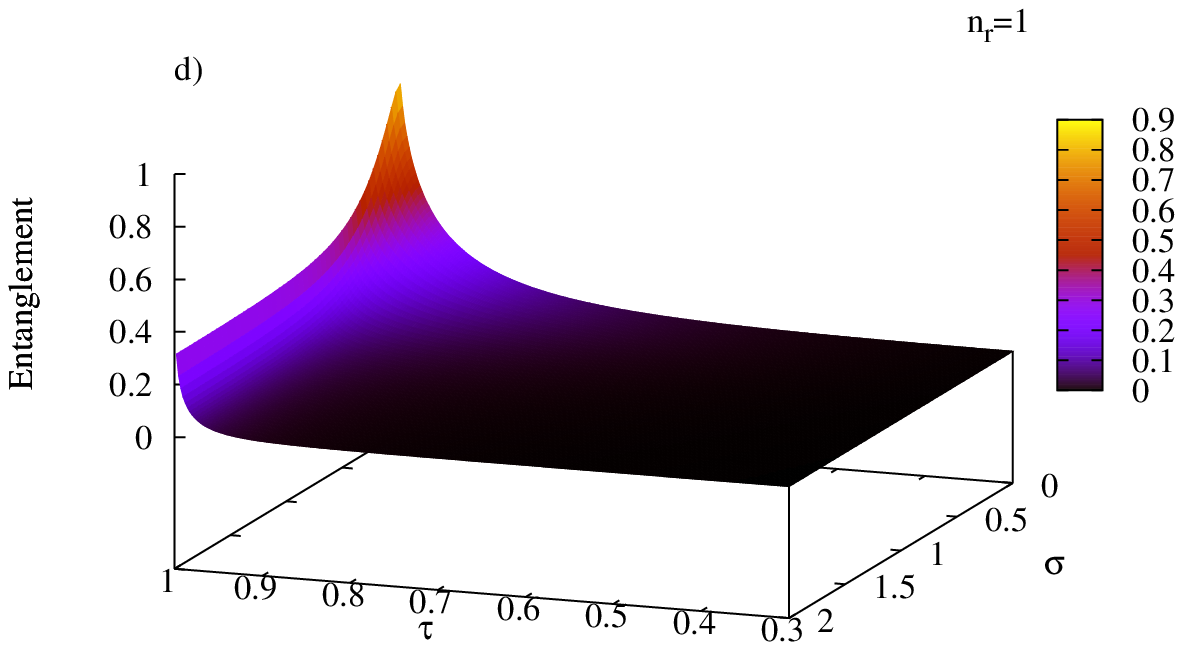}
\caption{(Color online) Entanglement of the first excited state in $n_R$ and $n_r$ of the three dimensional Moshinsky atom with two interacting electrons and magnetic field. a) Attractive interaction, antiparallel spin, b) Repulsive interaction, antiparallel spin, c) Attractive interaction, parallel spins, d) Repulsive interaction, parallel spins. All depicted quantities are dimensionless. \label{figure_6}}
\end{center}
\end{figure}
The behaviour of the entanglement exhibited by these states is shown in Figure 5
(for excited states with $\nu_r=1$ and $\nu_R=1$) and in Figure 6
(for excited states with $n_r=1$ and $n_R=1$).

It transpires from the calculations summarized in Figures 3-6 that
the amount of entanglement exhibited by the eigenstates of the
Moshinsky atom tends to decrease with the strength of the magnetic
field. To understand the physics behind this trend let us first
recall the general way in which entanglement depends on the
strength of the interaction between the two particles constituting
the system. Entanglement tends to increase with the relative strength of
the interaction. However, it is important to stress that the
determining factor here is not the ``absolute'' strength of the
interaction, but its strength as compared with the strength of the
external confining potential. In other words, entanglement
increases both if one increases the strength of the interaction
keeping constant the external potential or, alternatively, if the
strength of the confining potential is weakened while keeping
constant the interaction. These general trends have been observed
in all the atomic models where entanglement has been studied in
detail: the Moshinsky model, the Crandall model, and also in
Helium and in Helium-like atomic systems  \cite{YPD,MPDK10}. For
instance, when one considers decreasing values of the nuclear
charge $Z$ in Helium-like systems (weakening the Coulombic
confining potential) the entanglement of the system's ground state
increases \cite{MPDK10}. These general patterns admit a clear and
intuitive physical interpretation. When the external confining
potential becomes physically dominant (as compared with the
interaction) the behaviour of the system resembles the behaviour
of a system of independent, non-interacting particles, and
entanglement tends to decrease. On the other hand, when the
interaction is dominant (as compared with the confining potential)
the system's behaviour departs from that of a system of
non-interacting particles and entanglement tends to increase.

The dependence of entanglement with the magnetic field can now be physically understood.
This dependence follows the same general patterns explained above. Indeed, the basic
fact about the magnetic field in the Moshinsky model (which of course is a common
external field acting on both particles) determining its effect upon entanglement
is the following:  increasing the strength of the magnetic field tends to increase
the confining effect of the combined external fields (that is, the harmonic external
field and the magnetic field). To illustrate this basic property let us briefly consider
the behaviour of a single particle (in $3D$-space) under the combined effects of the
external fields (harmonic field plus uniform magnetic field) involved in the Moshinsky
model that we study here. The probability density corresponding to the ground state of
the particle is,
\begin{equation} \label{confinao}
\rho(x,y,z) = \frac{\sqrt{\omega } \sqrt{b^2+\omega ^2}}{\pi ^{3/2}} e^{-z^2 \omega
-\left(x^2+y^2\right) \sqrt{b^2+\omega ^2}}.
\end{equation}
A direct way to study the dependence of the confinement of this particle
on the strength of the magnetic field is to compute the entropy of the
spatial probability density and determine its behaviour with the magnetic
field (decreasing values of the entropy correspond to increasing confinement).
The linear entropy, $S^{(L)} = 1 - Tr[\rho^2] = 1 - \int [\rho(\mathbf{r})]^2 d\mathbf{r}$,
and von Neumann entropy,
$ S^{(vN)} = -Tr[\rho \ln (\rho) ] = - \int \rho(\mathbf{r}) \ln[\rho(\mathbf{r})] d\mathbf{r}$,
of the probability density (\ref{confinao}) are given, respectively, by
$S_{|gs\rangle}^{(L)}(\omega,b)= 1-\frac{\sqrt{\omega } \sqrt{b^2+\omega ^2}}{2 \sqrt{2} \pi ^{3/2}}$
and $ S_{|gs\rangle}^{(vN)}(\omega,b)=\frac{1}{2} \Bigl[ 3 \left(1 + \ln \pi  \right)-
\ln \omega  -\ln \left( b^2+\omega ^2\right)\Bigr]$.
The entropies $S_{|gs\rangle}^{(L)}$ and $S_{|gs\rangle}^{(vN)}$
describing the spatial ``spreading'' of the probability
density associated with ground state wave function are plotted against the magnetic
field in Figure 7. It can be clearly appreciated that confinement increases with
the intensity of the magnetic field. The probability densities corresponding to the
excited states of this single-particle case, as well as the two-particle spatial
probability densities corresponding to the eigenfunctions of the Moshinsky system,
also become more confined when the magnetic field becomes more intense. As already
explained, this behaviour is consistent with the decrease of entanglement with
an increasing magnetic field.
\begin{center}
\begin{figure}[h]
\includegraphics[scale=0.9]{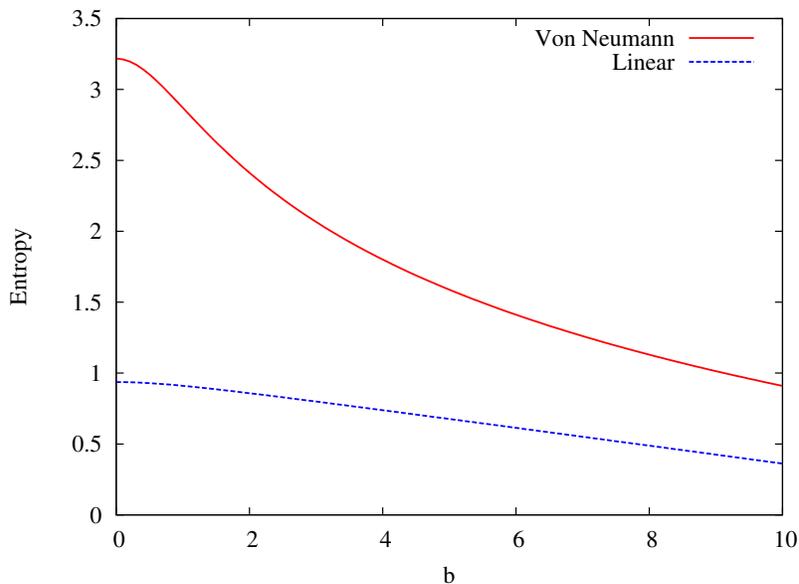}
\caption{Linear and von Neumann entropies as a function of $b$ with $\omega=1$}
 \end{figure}
\end{center}

\section{Perturbative Approach}

In this section we consider a perturbative approach to the two previously studied models, regarding the term that describe the interaction (between the three or two electrons) as a small perturbation ($\lambda^2 \sim 0$). Let us consider both systems governed by a Hamiltonian of the form

\begin{equation}
H=H_0+\lambda^2 H',
\end{equation}

\noindent where the unperturbed Hamiltonian $H_0$ takes different forms for each case (see Eqs. (\ref{H03e}) and (\ref{H02e})). $H_0$ corresponds to
three independent (non-interacting) particles and two independent particles in a magnetic field, respectively, and $\lambda^2 H'$ describes the interaction between the electrons, being $\lambda$  a small parameter. A perturbative treatment of this system involves an expansion of its eigenenergies and eigenstates in terms of powers of
$\lambda^2$. If this approach is valid we expect the gross
properties of the energy spectrum to be given by the eigenvalues
of the unperturbed Hamiltonian. It is clear that within this
scenario the leading, zeroeth-order contribution to the energy
spectrum is independent of the detailed structure of the
perturbation  $H'$. However, the situation is different
when, instead the energy, we calculate the entanglement of the
system's eigenstates. When the unperturbed energy eigenvalues are
degenerate the leading (zeroeth-order) contribution to the
eigenfunction's entanglement does depend, in general, on the
details of the perturbation.

Let us consider an $m$-fold degenerate energy level of $H_0$, with
an associated set of $m$ orthonormal eigenstates $|\psi_j\rangle,
\,\, j=1, \ldots m$. Since $H_0$ describes non-interacting
particles, the $m$ eigenstates $|\psi_j\rangle$ can always be
chosen to be Slater determinants written in terms of a family of
orthonormal single-particle states $|\phi^{(1,2,3)}_j \rangle$ in the case of three particles and in terms of $|\phi^{(1,2)}_j \rangle$ in the case of two particles. So we have for a three-particle system,

\[
|\psi_j\rangle=\frac{1}{\sqrt{6}}\left(|\phi^{(1)}_j\rangle |
\phi^{(2)}_j \rangle|\phi_j^{(3)}\rangle -
|\phi^{(1)}_j\rangle |\phi^{(3)}_j\rangle|\phi^{(2)}_j\rangle+|
\phi^{(2)}_j\rangle |\phi^{(3)}_j \rangle|\phi^{(1)}_j\rangle- \right.
\]
\[
 \left. |\phi^{(2)}_j\rangle |\phi_j^{(1)} \rangle|\phi_j^{(3)}\rangle
+|\phi^{(3)}_j\rangle |\phi^{(1)}_j \rangle|\phi_j^{(2)}\rangle-|
\phi^{(3)}_j\rangle |\phi^{(2)}_j \rangle|\phi_j^{(1)}\rangle \right),
\]
and for a two-particles systems
\[
 |\psi_j\rangle = \frac{1}{\sqrt{2}} \left(|\phi_j^{(1)}\rangle|
\phi_j^{(2)}\rangle-|\phi_j^{(2)}\rangle|\phi_j^{(1)}\rangle\right).
\]

All the members of the subspace ${\cal H}_s$ spanned by the states $|\psi_j\rangle$
are eigenstates of $H_0$ corresponding to the same eigenenergy.
The different members of this subspace have, in general, different
amounts of entanglement. Typically, the interaction $H'$ will lift
the degeneracy at least partially of the degenerate energy level.
If we solve the eigenvalue problem corresponding to the
(perturbed) Hamiltonian $H$ and take the limit $\lambda
\rightarrow 0$, the perturbation $H^{\prime}$ will ``choose'' one
particular basis $\{ |\psi'_k\rangle_{\lambda\to 0} \}$ among the
infinite possible basis of ${\cal H}_s$. The states constituting
this special basis will in general be entangled. These states are
of the form \cite{Desai10}

\begin{equation}
|\psi'_k\rangle_{\lambda\to 0}=\sum_{j=1}^m c_{kj}|\psi_j\rangle,
\end{equation}

\noindent where the $m$-dimensional vectors $v_k^T=(c_{k1},...,c_{km})$
are the eigenvectors of the $m\times m$ $\tilde H$ matrix with elements given by,

\begin{equation}\label{matrix-elements}
\tilde H_{ij}=\langle\psi_i|H'|\psi_j\rangle.
\end{equation}

\noindent
It is then clear that, in the limit $\lambda \rightarrow 0$
the eigenstates of $H$ will in general be entangled.

Let $\tilde m$ be the number of different single-particle states
within the family $\{|\phi^{(1,2,3)}_j \rangle, \, 1, \ldots, m
\}$ or $\{|\phi^{(1,2)}_j \rangle, \, 1, \ldots, m
\}$. $\tilde m$ tends to increase with $m$ which, in turn, tends
to increase with energy; that is, $\tilde m$ tends to increase as
one considers higher excited states. This explains (at least in
part) why the range of entanglement values available to the
eigenstates $\{ |\psi'_k\rangle_{\lambda\to 0} \}$ tends to
increase with energy. Indeed, the amount of entanglement
that can be achieved for a given energy of a $N$-fermion system admits an upper bound given by

\begin{equation} \label{upSL}
\varepsilon_{SL} =1-\frac{N}{\tilde m},
\end{equation}

\subsection{Moshinsky model with three electrons}

Let the unperturbed Hamiltonian be,

\begin{equation}\label{H03e}
H_0=-\frac{1}{2}\frac{\partial^2}{\partial
x_1^2}-\frac{1}{2}\frac{\partial^2}{\partial
x_2^2}-\frac{1}{2}\frac{\partial^2}{\partial
x_3^2}+\frac{1}{2}\omega^2x_1^2
+\frac{1}{2}\omega^2x_2^2+\frac{1}{2}\omega^2x_3^2
\end{equation}

\noindent and the perturbation,

\begin{equation}
\lambda^2 H'=\lambda^2 \frac{1}{2}[(x_1-x_2)^2+(x_2-x_3)^2+(x_1-x_3)^2].
\end{equation}
Then, we have $H=H_0+\lambda^2 H'$. When $\lambda=0$ the model consists of three-independent harmonic oscillators with the same natural frequency. Let $|n \pm\rangle$ ($n=0,1,2...$) the eigenstate of each of these oscillators. For the first excited state which is four-fold degenerate, let $\{|0,\pm\rangle,|1,\pm\rangle,|2,\pm \rangle\}$ be the single-particle orthonormal basis. Then, for $\lambda=0$, we can choose the four eigenstates with zero entanglement, all of them with the same energy as  $|011\rangle_{R_1R_2R_3}$ and $|110\rangle_{R_1R_2R_3}$

\begin{eqnarray}\label{eigenvec4}
|\psi_1\rangle&=&|0+,0-,2+|\nonumber\\
|\psi_2\rangle&=&|0+,0-,2-|\nonumber\\
|\psi_3\rangle&=&|0+,1+,1-|\nonumber\\
|\psi_4\rangle&=&|0-,1+,1-|\nonumber\\
\end{eqnarray}
where we have introduced the notation
\[
 |i,j,k|= \frac{1}{\sqrt{6}} \left( |i,j,k\rangle-|i,k,j\rangle+|j,k,i
\rangle-|j,i,k\rangle+|k,i,j\rangle-|k,j,i\rangle \right)
\]
and $i=j=k=0\pm,1\pm,2\pm $.

For the first excited energy level of $H_0$ ($E=\frac{7}{2}\omega$), we have

\begin{equation}
\tilde H\propto\begin{pmatrix}
  4 & 0 & \frac{1}{\sqrt{2}} & 0 \\
  0 & 4 & 0 & \frac{1}{\sqrt{2}} \\
  \frac{1}{\sqrt{2}} & 0 & \frac{7}{2} & 0 \\
  0 & \frac{1}{\sqrt{2}} & 0 & \frac{7}{2}
\end{pmatrix},
\end{equation}

\noindent and the corresponding eigenvectors can be written as

\begin{eqnarray}\label{eigenvec4prima}
|\psi_1'\rangle&=&\sqrt{\frac{2}{3}}(-\frac{1}{\sqrt{2}}|\psi_2\rangle+|\psi_4\rangle)\nonumber\\
|\psi_2'\rangle&=&\sqrt{\frac{2}{3}}(-\frac{1}{\sqrt{2}}|\psi_1\rangle+|\psi_3\rangle)\nonumber\\
|\psi_3'\rangle&=&\frac{1}{\sqrt{3}}(\sqrt{2}|\psi_2\rangle+|\psi_4\rangle)\nonumber\\
|\psi_4'\rangle&=&\frac{1}{\sqrt{3}}(\sqrt{2}|\psi_1\rangle+|\psi_3\rangle).
\end{eqnarray}

\noindent In the decoupled limit the eigenstates $|011\rangle_{R_1R_2R_3}$ and
$|110\rangle_{R_1R_2R_3}$ tend to the states \eqref{eigenvec4prima}
(or any combination of them) which have $ \varepsilon=\frac{8}{27}$.
This value coincides with the amount of entanglement for the first excited
state obtained from the exact calculation in the limit $\tau\to 0$
(equivalently $\lambda\to 0$).

The states $| \psi_1' \rangle $ and $|\psi_2' \rangle $
($|\psi_3' \rangle $ and $|\psi_4' \rangle $) share the same energy
eigenvalue. A linear combination of eigenstates  sharing the
same eigenenergy is also a valid energy eigenstate. Then,
let us consider for instance $|\psi_{34}' \rangle =
\cos \theta \, |\psi_3' \rangle + \sin \, \theta |\psi_4' \rangle $,
($0\leq \theta < 2\pi$).  As already discussed at the end of
Subsection III.B, the amount of entanglement of these linear
combinations does not depend on $\theta$.

We have seen that in the case of some excited states of the
Moshinsky model an arbitrarily weak interaction between the
particles leads to a finite amount of entanglement. This naturally
suggests the following issues: to what extent is this weak-interaction
entanglement robust? What happens with this entanglement if
some other small perturbation acts upon the system? The detailed
entanglement features of the eigenstates corresponding to this
scenario will evidently depend on the precise form of the new perturbation.
Therefore, these entanglement properties can only be studied in
a case-by-case way. However, it is possible to gain some valuable
insights on the robustness of the weak-interaction entanglement
by recourse to a statistical approach. We can consider the typical
features of the weak-interaction entanglement corresponding to a
random perturbation.

Let us consider again the four eigenstates (\ref{eigenvec4}) of the
unperturbed (with no interaction) system. Any weak perturbation
acting on top of the already considered weak interaction will lead
(in the lowest order of perturbation theory for a degenerate
eigenenergy) to a new set of four perturbed energy eigenstates
that will be linear combinations of the four unperturbed states
(\ref{eigenvec4}). That is, the new perturbed states are orthonormal
states belonging to the four-dimensional linear space spanned by
the states (\ref{eigenvec4}). We can consider the statistical
distribution of entanglement values corresponding to random states
in this subspace uniformly distributed according to the
Haar measure (see \cite{BCPP02,HIPZ10} and references therein).
To this end we generate three-electron states randomly distributed 
according to the Haar measure of the form,

\begin{equation}
 |\psi'\rangle=\sum_{i=1}^4c_i|\psi_i\rangle,
\end{equation}

\noindent with $|\psi_i\rangle,\,\, i=1,\ldots, 4$ as given in Eq.
(\ref{eigenvec4}). A state of the form $|\psi'\rangle$ can be thought
as an eigenvector corresponding to an arbitrary perturbation.
The amount of entanglement of the state $|\psi'\rangle$ is,

\begin{equation}\label{gen-ent}
 \varepsilon(|\psi'\rangle)=1-\frac{1}{3}\left[2\left(|c_1|^2+|c_2|^2\right)^2+2\left(|c_3|^2+|c_4|^2\right)^2+1\right].
\end{equation}

\noindent
Optimizing Eq. (\ref{gen-ent}) we obtain the maximum possible value of entanglement
$\varepsilon_m(|\psi'\rangle)=\frac{1}{3}$ associated to the state with coefficients
satisfying: $|c_1|^2+|c_2|^2=|c_3|^2+|c_4|^2=\frac{1}{2}$.

In Table \ref{T-PDF} we show the percentual number of
three-electron pure states belonging to the linear subspace
spanned by the states (\ref{eigenvec4}) that have entanglement
values in different ranges. To compute these percentual values
we generated $10^7$ random states distributed according to
the Haar measure. The average and maximum entanglement values
corresponding to perturbed states spanned by (\ref{eigenvec4})
are also given in Table \ref{T-PDF}. These results constitute
suggestive evidence for the robustness of the entanglement
exhibited by excited eigenstates of the Moshinsky atom in the
weak-interaction limit. Indeed, the statistical study summarized
in Table \ref{T-PDF} suggests that any new perturbation
is likely to produce a small decrease in the entanglement of
the excited state considered here ($\varepsilon=8/27 \approx 0.2963$)
but, in the typical case, it will still result in an appreciable
amount of entanglement in the weak-interaction limit.

\begin{table}[h]
 \begin{center}
\begin{tabular}{|c|c|c|}
  \hline
  $\%$ of states in & $0<\varepsilon\le 1/9$ &  $ 4.75\%$ \\ \cline{2-3}
  different  & $1/9<\varepsilon\le 2/9$  &  $18.25\%$ \\ \cline{2-3}
  entanglement ranges & $2/9<\varepsilon \le 1/3 $ & $77\%$  \\
  \hline
 average entanglement & $\langle\varepsilon\rangle=0.26667$ & \\
  \hline
  maximum entanglement & $\varepsilon_m=1/3$ & \\
   \hline
\end{tabular}
\caption{Entanglement distribution for perturbed 
excited states of a three-electron Moshinsky 
system. These states correspond to a four-fold 
degenerate unperturbed energy level.}\label{T-PDF}
\end{center}
\end{table}

Let us now study the entanglement properties for the second excited state of the
Moshinsky atom. In this case, we have ten-fold degenerate eigenstates,
(all of them with $E=\frac{9}{2}\omega$). The single-particle orthonormal basis
is given by $\{|0,\pm\rangle,|1,\pm\rangle,|2,\pm\rangle,|3,\pm\rangle\}$. The
matrix of the harmonic perturbation can be expressed as follow:

\begin{equation}\tilde H\propto\begin{pmatrix}
  \frac{9}{2} & 0 & 0 & 0 & -\frac{\sqrt{3}}{2} & 0 & \frac{\sqrt{3}}{2} & 0 & 0 & 0 \\
  0 & \frac{9}{2} & 0 & 0 & 0 & -\frac{\sqrt{3}}{2} & 0 & \frac{\sqrt{3}}{2} & 0 & 0 \\
  0 & 0 & 6 & 0 & 0 & 0 & 0 & 0 & 0 & 0\\
  0 & 0 & 0 & 5 & 1 & 0 & 0 & 0 & 0 & 0 \\
  -\frac{\sqrt{3}}{2} & 0 & 0 & 1 & \frac{9}{2} & 0 & \frac{1}{2} & 0 & 0 & 0\\
  0 & -\frac{\sqrt{3}}{2} & 0 & 0 & 0 & \frac{11}{2} & 0 & \frac{1}{2} & 0 & 0 \\
 \frac{\sqrt{3}}{2} & 0 & 0 & 0 & \frac{1}{2} & 0 & \frac{11}{2} & 0 & 0 & 0\\
  0 & \frac{\sqrt{3}}{2} & 0 & 0 & 0 & \frac{1}{2} & 0 & \frac{9}{2} & 1 & 0\\
  0 & 0 & 0 & 0 & 0 & 0 & 0 & 1 & 5 & 0 \\
  0 & 0 & 0 & 0 & 0 & 0 & 0 & 0 & 0 & 6
\end{pmatrix}.
\end{equation}

Following a similar procedure and after a laborious algebra we compute the entanglement amount of the eigenvectors of $\tilde H$,

\begin{eqnarray}
\varepsilon(|\psi_1'\rangle)=\varepsilon(|\psi_6'\rangle)=0\nonumber\\
\varepsilon(|\psi_7'\rangle)=\varepsilon(|\psi_8'\rangle)=\varepsilon(|\psi_9'\rangle)=\varepsilon(|\psi_{10}'\rangle)=\frac{4}{9}\nonumber\\
\varepsilon(|\psi_3'\rangle)=\varepsilon(|\psi_4'\rangle)=\frac{1}{4}\nonumber\\
\varepsilon(|\psi_2'\rangle)=\varepsilon(|\psi_5'\rangle)=\frac{20}{49}
\end{eqnarray}
where $|\psi'_j \rangle$ ($j=1,...,10$) are the eigenvectors of the $\tilde H$ matrix. The states $|\psi_j'\rangle$ with $j=1,...,6$ share the same eigenvalue. The same occurs for the state pairs $|\psi'_7\rangle$ and $|\psi'_8\rangle$, and $|\psi'_9\rangle$ and $|\psi'_{10}\rangle$. As we mention before, the interaction lift only partially the degeneracy. The degeneracy due to the spin degree of freedom ($S_z=\pm\frac{1}{2}$) is present in all the states. The obtained values agree with some of those calculated in section 3. Different combinations of the states sharing eigenvalues result in the non-coincident entanglement amount $\varepsilon_{021}=\frac{43}{108}$ For instance, let $|\psi_{56}'\rangle=p\psi_5'+\sqrt{1-p^2}\psi_6'$, ($0\leq p\leq 1$); then

\begin{equation}
 \varepsilon(p)=\frac{4}{147} p^2 (8 p^2 + 7),
\end{equation}

\noindent and $\varepsilon_{021}$ is re-obtained for $p\sim 0.992$.

\subsection{Moshinsky model with two electrons in a uniform magnetic field}

We consider also a perturbative approach for a  three-dimensional Moshinsky atom with two electrons in a magnetic field. Let the unperturbed Hamiltonian be,

\begin{equation}\label{H02e}
 H_0 = \frac{1}{2} (p_1^2+p_2^2) + \frac{\omega^2}{2} (r_1^2+r_2^2) +
\frac{b^2}{2} (x_1^2+y_1^2+x_2^2+y_2^2)+ b (L_{1z}+L_{2z})
\end{equation}

\noindent and the perturbation,

\begin{equation}
\lambda^2 H'=\frac{\lambda^2}{2} (\mathbf{r}_1-\mathbf{r}_2)^2.
\end{equation}

\noindent The eigenenergies of $H_0$ are given by Eq. \eqref{totalenergy}, taking $\nu_R=\nu_1$, $\nu_r=\nu_2$, $m_R=m_1$, $m_r=m_2$, $n_R=n_1$, $n_r=n_2$ and setting $\lambda=0$. Then, for the excited states of $H_0$ with energy given by

$$E_{\nu m}=\omega\left(2y+\frac{2}{y}+1\right), \,\,\,\,\, y=\left(1+\frac{b^2}{\omega^2}\right)^{\frac{1}{2}}+\frac{b}{\omega},$$

\noindent resulting of setting one of the quantum numbers $\nu_{1}$, $\nu_{2}$, $|m_{1}|$, $|m_{2}|$ equal to one and the rest equal to zero, we obtain

\begin{equation}
\tilde H\propto\begin{pmatrix}
  c_1 & 0 & 0 & 0 & 0 & 0 & 0 & 0   \nonumber\\
  0 & c_1 & 0 & 0 & 0 & c_2 & -c_2 & 0 \nonumber\\
  0 & 0 & c_1 & 0 & 0 & -c_2 & c_2 & 0 \nonumber\\
  0 & 0 & 0 & c_1 & 0 & 0 & 0 & 0  \nonumber\\
  0 & 0 & 0 & 0 & c_1 & 0 & 0 & 0 \nonumber\\
  0 & c_2 & -c_2 & 0 & 0 & c_1 & 0 & 0\nonumber\\
  0 & -c_2 & c_2 & 0 & 0 & 0 & c_1 & 0  \nonumber\\
  0 & 0 & 0 & 0 & 0 & 0 & 0 & c_1
\end{pmatrix},
 \end{equation}

\noindent where $c_1=\frac{1}{2\omega}+\frac{2}{\sqrt{b^2+\omega^2}}$ and $c_2$ is obtained numerically and its exact numerical value is not relevant for the next calculations.

$\tilde H$ has six degenerate eigenvectors and two non-degenerate ones that take the following entanglement values: $\{0,\frac{1}{2},\frac{3}{4}\}$. The entanglement value obtained from the exact computations in the limit $\lambda\to 0$ of the states with the same energy, $|100,000\rangle_{Rr}$ and $|000,100\rangle_{Rr}$, coincide with one of the above values ($\varepsilon=\frac{3}{4}$).

We consider also states setting $n_{1}=1$ or $n_{2}=1$ and the rest equal to zero, with energy given by $$E_{n}=\omega\left(y+\frac{1}{y}+2\right)$$ and with $y$ as before. For these excited states we obtained

\begin{equation}
\tilde H\propto\begin{pmatrix}
  d_1+d_2 & 0 & 0 & 0    \nonumber\\
  0 & d_1 & d_2 & 0 \nonumber\\
  0 & d_2 & d_1 & 0 \nonumber\\
  0 & 0 & 0 & d_1+d_2
\end{pmatrix},
 \end{equation}

\noindent where $d_1=\frac{1}{\omega}+\frac{1}{\sqrt{b^2+\omega^2}}$ and $d_2=\frac{1}{2\omega}$. This matrix has three eigenvectors corresponding to the same eigenvalues and with entanglement $\{0,\frac{1}{2}\}$ and one non-degenerate eingevector with entanglement $\frac{1}{2}$. Again the obtained results are in perfect accordance with the entanglement obtained for  states $|001,000\rangle_{Rr}$ and $|000,001\rangle_{Rr}$ in the decoupled regime.

\section{Conclusions}

We explored the entanglement properties of two versions of the Moshinsky model: one comprising
three electrons and another one consisting of two electrons in a uniform external magnetic field.
The amount of entanglement of the eigenstates of the three-electron system considered here depends
only on the dimensionless parameter $\tau$ describing the relative strength of the interaction
between the particles (as compared with the strength of the external confining potential).
We obtained closed analytical expressions for the amount of entanglement of the ground, first and
second excited states. As a general trend we found that the entanglement exhibited for these states
tends to increase both with the state's energy and with the strength of the interaction between
the particles (that is, with $\tau$). Non-vanishing entanglement is obtained in the limit of
 vanishing interaction in the case of excited states. This (apparent) discontinuous behaviour
of the entanglement is related to the degeneracy of the energy levels of the ``unperturbed''
 Hamiltonian describing non-interacting particles. The non-vanishing entanglement in the limit of
zero interaction is determined by the particular basis of $H_0$ ``chosen'' by the interaction.
We also found that in the case of an attractive interaction the eigenstates' entanglement
approaches its maximum possible value in the limit of an infinitely large interaction. On
the other hand, in the case of a repulsive interaction the maximum possible entanglement is
obtained when the interaction strength approaches a finite, critical limit  value corresponding
to $\tau_c = \frac{1}{\sqrt{3}}$. The system does not admit bound eigenstates when the strength
of the (repulsive) interaction is equal or larger than the one corresponding to $\tau_c$.

As far as the entanglement's dependence on the interaction
strength and the energy are concerned, the behavior of the
Moshinsky model with two electrons in a uniform magnetic field is
similar to the one observed in the three-electron model. With
regards to the external magnetic field, we found that the
eigenstates' entanglement decreases when considering increasing
magnetic fields. In the limit of very strong magnetic fields the
entanglement approaches a finite asymptotic value that depends on
the interaction strength. The essential aspect of the magnetic
field in the Moshinsky model that determines its effect upon the
amount of entanglement exhibited by the system's eigenstates is
the following: increasing the intensity of the magnetic field
tends to increase the confining effect of the combined external
fields (that is, the harmonic external field and the magnetic
field). For a given strength of the interaction between the
particles, this increasing confinement leads (according to a
general pattern that has been observed in all atomic models where
entanglement was studied in detail) to a decrease in the
eigenstates' entanglement. As happens in the case of the
three-electron model, a perturbative treatment highlights the
essential role played by the degeneracy of the energy levels of
the interactionless system in determining how the eigenstates'
entanglement depends on the interaction strength and on the
energy.

\acknowledgments This work was partially supported by the Projects FQM-2445 and FQM-4643 of the
Junta de Andalucia and the grant FIS2011-24540 of the Ministerio de Innovaci\'on y Ciencia, Spain.
We belong to the Andalusian research groups FQM-207 and FQM-020. A.P.M. acknowledges support by
GENIL through YTR-GENIL Program.


\end{document}